\documentclass[pre,amssymb,amsmath,twocolumn,aps,showpacs,superscriptaddress]{revtex4}
\usepackage{amsmath,graphicx}

\usepackage[dvips]{epsfig}
\usepackage[usenames,dvipsnames]{color}
\usepackage[caption=false]{subfig}
\usepackage{verbatim}
\newcommand{\nmax}{n_{\mathrm{max}}} 
\newcommand{\peq}{p^{\mathrm{eq}}_n(\mu)}
\newcommand{\nav}{\langle n \rangle}
\newcommand{\nvar}{\langle n^2 \rangle - \langle n \rangle^2}

\begin{document}
\title{Diffusion of interacting particles in discrete geometries: Equilibrium and dynamical properties}
\author{T. Becker}
\email{thijsbecker@gmail.com}
\affiliation{Hasselt University, B-3590 Diepenbeek, Belgium}
\author{K. Nelissen}
\affiliation{Departement Fysica, Universiteit Antwerpen, Groenenborgerlaan 171, B-2020 Antwerpen, Belgium}
\affiliation{Hasselt University, B-3590 Diepenbeek, Belgium}
\author{B. Cleuren}
\affiliation{Hasselt University, B-3590 Diepenbeek, Belgium}
\author{B. Partoens}
\affiliation{Departement Fysica, Universiteit Antwerpen, Groenenborgerlaan 171, B-2020 Antwerpen, Belgium}
\author{C. Van den Broeck}
\affiliation{Hasselt University, B-3590 Diepenbeek, Belgium}
\date{\today}
\begin{abstract}
We expand on a recent study of a lattice model of interacting particles [Phys.~Rev.~Lett.~111, 110601 (2013)]. The adsorption isotherm and equilibrium fluctuations in particle number are discussed as a function of the interaction. Their behavior is similar to that of interacting particles in porous materials. 
Different expressions for the particle jump rates are derived from transition-state theory. Which expression should be used depends on the strength of the interparticle interactions.
Analytical expressions for the self- and transport diffusion are derived when correlations, caused by memory effects in the environment, are neglected. The diffusive behavior is studied numerically with kinetic Monte Carlo (kMC) simulations, which reproduces the diffusion including correlations. The effect of correlations is studied by comparing the analytical expressions with the kMC simulations. It is found that the Maxwell-Stefan diffusion can exceed the self-diffusion. To our knowledge, this is the first time this is observed.
The diffusive behavior in one-dimensional and higher dimensional systems is qualitatively the same, with the effect of correlations decreasing for increasing dimension. The length dependence of both the self- and transport diffusion is studied for one-dimensional systems. For long lengths the self-diffusion shows a $1/L$ dependence. Finally, we discuss when agreement with experiments and simulations can be expected. The assumption that particles in different cavities do not interact is expected to hold quantitatively at low and medium particle concentrations if the particles are not strongly interacting. 
\end{abstract}
\pacs{05.40.Jc, 02.50.--r, 05.60.Cd, 66.30.Pa}
\maketitle
\section{\label{sec::intro}Introduction}

Nanoporous materials such as zeolites and metal-organic frameworks (MOFs) are molecular-scale structures consisting of interconnected channels or cavities. Particles diffusing in these materials are tightly confined, leading to diffusive behavior that can markedly differ from bulk diffusion \cite{book_diffnano}. Due to their large surface area and molecular-scale structure they are ideally suited for applications such as catalysis \cite{satterfield1981mass}, particle separation \cite{CRSmit}, and carbon dioxide storage \cite{sumida2011carbon}. Developments in the synthesis of new porous materials \cite{Natureyaghi2003,porousapp} have yielded a large increase in available materials, which could allow for a fine-tuning of the properties of the material, depending on the application.

New experimental techniques give a detailed look of particle diffusion in nanoporous materials \cite{Naturekarger2014}. A theoretical understanding has been achieved using different approaches. Molecular dynamics (MD) simulations \cite{CRSmit,CSRkrishna2012,PRLbeerdsen2006} can incorporate the full atomic structure of the material in the simulations.However, they can be computationally time-consuming. 
Because many nanoporous materials consist of discrete sorption sites or cavities separated by narrow windows, particle diffusion often can be described by a hopping process on a lattice \cite{saravanan1997modeling,saravanan1997modeling_2,JCPBdemontis2008,IRPCauerbach,SurfScireed1981,bhide1999dependence}. Another popular approach is continuum models where the particles follow a stochastic dynamics \cite{zwanzig1992diffusion,Burada2009,PREcarvalho}, which has been used to study phenomena such as stochastic resonance \cite{ghosh2010geometric} and novel particle separation methods \cite{reguera2012entropic}.

Diffusion in equilibrium is characterized by the self-diffusion coefficient. It describes the average mean-squared displacement (MSD) of a single particle. The transport diffusion coefficient, on the other hand, describes the magnitude of the particle flux in response to a concentration gradient. It is therefore the relevant quantity in nonequilibrium conditions. In general, these two diffusion coefficients differ because of particle interactions. Understanding their concentration dependence and relation as a function of the particle interaction is of both fundamental and practical interest. Recently we introduced a lattice model \cite{PRLbecker} which describes the diffusive hopping of particles in a compartmentalized system. All interactions are defined by a single function, namely the equilibrium free energy of the particles in a compartment (also called a \textit{cavity}). Using this model, we were able to provide a simple interpretation of an experiment of methanol and ethanol diffusion in the nanoporous material MOF zeolitic imidazolate framework 8 (ZIF-8) \cite{PRLchmelik2010}. In contrast to previous experiments, the self-diffusion was found to exceed the transport diffusion at certain particle concentrations. From molecular dynamics simulations it was shown that this was the result of particle clustering \cite{LANGMUIRkrishna2010_1}. Because in our model particle clustering is connected to the equilibrium free energy in a straightforward way, the effect of clustering on the diffusion could be understood from a simple analytical argument. Combined with numerical simulations, it was shown that in our model particle clustering is a necessary condition for the self-diffusion to exceed the transport diffusion.

We present the following results. Particle clustering in porous materials has a distinct effect on the adsorption isotherm and the equilibrium fluctuations in particle number \cite{LANGMUIRkrishna2010_1,LANGMUIRkrishnaHbond,LANGMUIRkrishnaTc}. We discuss how this can be understood from the shape of the equilibrium free energy in our model. Particles jump between the cavities with rates that depend on the interaction. Different forms of these jump rates are calculated from transition-state theory (TST). It is explained which rates should be applied for different kinds of systems. Analytical expressions for the self- and transport diffusion were derived in Ref.~\cite{PRLbecker} for a system of length 1. We show that these are the expressions for the diffusion coefficients if one ignores all dynamical correlations. The diffusive behavior is investigated for different jump rates and interactions. The effect of correlations is discussed in detail. In Ref.~\cite{PRLbecker} a quantitative agreement with the experimental results from Ref.~\cite{PRLchmelik2010} was found. We discuss here the quality of the assumptions made in our model, and when agreement with experiments can be expected. 

The paper is organized as follows. In Sec.~\ref{sec::model} we introduce the model. The necessary concepts of diffusion theory are given in Sec.~\ref{sec::diffth}. Section \ref{sec::eqprop} discusses the equilibrium properties of the model. The distribution of particle occupation in the cavities is investigated as a function of the interaction. The behavior of the adsorption isotherms as a function of both the interaction and the confinement (maximum number of particles in each cavity) is studied. In Sec.~\ref{sec::dynamic} we present an analysis of the dynamical properties of the model. Possible forms of the transition rates are calculated in Sec.~\ref{sec::transrates}. The analytical expressions for the self- and transport diffusion when ignoring correlations are derived in Sec.~\ref{sec::DMFA}. The numerical simulations that are used to determine the self- and transport diffusion are explained in Sec.~\ref{sec::numsim}. We discuss the diffusive behavior for different interactions and rates in Sec.~\ref{sec::diffandcorr}. In Sec.~\ref{sec::assumptions} we discuss when agreement with experiments and simulations can be expected. A conclusion is presented in Sec.~\ref{sec::conclude}.
\section{\label{sec::model}The Model}
\begin{figure}
\centering
\includegraphics[width=0.9\columnwidth]{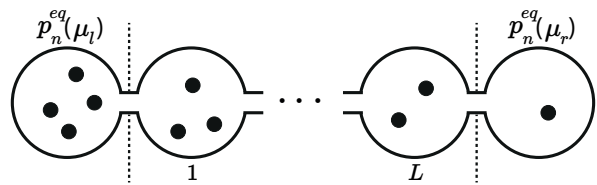}
\caption{The system, shown here between dashed lines, consists of an array of cavities connected by narrow passages. On the boundaries it is connected to uncorrelated cavities with the equilibrium distribution.}
\label{fig::generalsystem}
\end{figure}

The materials we consider consist of a large array of cavities, which are connected to each other by narrow passages, also called windows; see Fig.~\ref{fig::generalsystem}. In such a setup, it is natural to assume that the time spent by a particle in a cavity before moving to one of its neighbors is much larger than the equilibration time of particles inside a cavity. This allows us to coarse grain the intracavity degrees of freedom \cite{PREesposito2012}. Interactions are described by the equilibrium free energy $F(n)$, depending only on the number of particles $n$ in the cavity. Contributions to $F(n)$ are the result of particle-particle and particle-wall interactions inside a cavity. When the system is in equilibrium with a particle reservoir at chemical potential $\mu$ and temperature $T$, the probability to have $n$ particles in any cavity is equal to
\begin{equation}\label{eq::probeq}
p^{\mathrm{eq}}_n(\mu) = \left[ \mathcal{Z}(\mu)\right]^{-1} e^{- \beta \left[F(n)-\mu n\right]},
\end{equation}
with $\beta=(k T)^{-1}$, $k$ the Boltzmann constant, and \( {\mathcal{Z}} \) the grand-canonical partition function:
\begin{equation}
{\mathcal{Z}}(\mu) = \sum_{n=0}^{\nmax} e^{- \beta \left[F(n)-\mu n\right]}.
\end{equation}
Averages over the equilibrium distribution Eq.~\eqref{eq::probeq} are denoted by $\langle \cdot \rangle$, e.g.,
\begin{equation}\label{eq::defnav}
\nav (\mu) = \sum_{n=0}^{\nmax} n \peq.
\end{equation}
Since the equilibrium distribution is known for any given $F(n)$ and $\mu$, all equilibrium quantities can be calculated analytically in function of these two variables.
For later reference, we introduce the grand potential $\Omega(n,\mu) = F(n)-\mu n$, which captures the $n$ dependence of the probability. Confinement limits the amount of particles in a cavity and is represented in our model by $n_{\mathrm{max}}$, which is the maximal number of particles a cavity can contain.

A schematic representation of the model in one dimension is given in Fig.~\ref{fig::generalsystem}. It consists of pairwise connected cavities numbered from 1 to $L$. Because we integrate out the intracavity degrees of freedom we can identify the cavities with sites on a lattice. The center-to-center distance between two cavities is equal to $\lambda$. A particle jumps from a cavity containing $n$ particles to a cavity containing $m$ particles with probability per unit time $k_{nm}$. These rates satisfy local detailed balance:
\begin{align}\label{eq::localdetbalance}
\frac{k_{n m}}{k_{m+1,n-1}} &= \frac{p^{\mathrm{eq}}_{n-1}(\mu) p^{\mathrm{eq}}_{m+1}(\mu)}{p^{\mathrm{eq}}_n(\mu) p^{\mathrm{eq}}_m(\mu)} \\
&= e^{- \beta \left[F(n-1) + F(m+1) - F(n) - F(m)  \right]}.
\end{align}
This ensures that, when the system is in equilibrium, there are no net currents and that the probability distribution equals the equilibrium distribution Eq.~\eqref{eq::probeq}.
Particles can enter or leave the system through the boundaries, which are connected to (particle) reservoirs. The left and right reservoirs have, respectively, chemical potential $\mu_l$ and $\mu_r$. A reservoir is modeled as a cavity characterized by the equilibrium distribution $\peq$ ($\mu$ is either $\mu_l$ or $\mu_r$), whose state is uncorrelated from the cavity it is connected to. The rates at which a reservoir cavity at chemical potential $\mu$ adds ($k^+_n$) or removes ($k^-_n$) one particle from a cavity containing $n$ particles are
\begin{equation}\label{eq::boundaryrates}  
k^+_n = \sum_{m=1}^{\nmax} k_{mn} p^{\mathrm{eq}}_m(\mu) ; \quad k^-_n = \sum_{m=0}^{\nmax-1} k_{nm} p^{\mathrm{eq}}_m(\mu).
\end{equation}
\section{\label{sec::diffth}Diffusion Theory}

In this section we present the necessary theory that will be used later. For simplicity, we assume that the diffusion is isotropic. The average particle concentration at position $\mathbf{r}$ is denoted by $c = c(\mathbf{r})$.

\begin{figure}
\centering
\includegraphics[width=0.9\columnwidth]{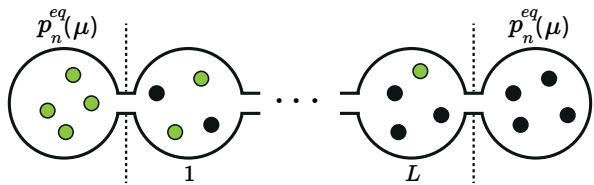}
\caption{(Color online) Measuring the self-diffusion: A concentration gradient of labeled particles (open green circles) is introduced under overall equilibrium conditions.}
\label{fig::generalsystem_label}
\end{figure}

The self-diffusion coefficient $D_s$ describes the average MSD of a single particle in a system at equilibrium, in the long-time limit: 
\begin{equation}\label{eq::selfdiff}
D_s = \lim_{t\rightarrow \infty} \frac{1}{2 d t} \overline{ \left[ \mathbf{r}(t) - \mathbf{r}(0) \right]^2 }= \lim_{t\rightarrow \infty} \frac{1}{2 d t} \overline{ \Delta \mathbf{r}^2(t) },
\end{equation}
where $d$ is the dimension of the system, $\mathbf{r}$ the position of the particle, and the overline denotes the average over all equilibrium trajectories. A common way of measuring this coefficient is by labeling a subset of the particles in the system (denoted by $*$); see Fig.~\ref{fig::generalsystem_label}. Particles in the reservoir cavities are labeled with different percentages, resulting in a concentration gradient $\nabla c^*$ of labeled particles under overall equilibrium conditions. The resulting flux $\mathbf{j}^*$ of  the labeled particles reads \cite{book_diffnano}:
\begin{equation}\label{eq::selfdiff2}
\mathbf{j}^* = - D_s \nabla c^*.
\end{equation}

The transport diffusion coefficient $D_t$, also called Fick or chemical diffusion, quantifies the particle flux $\mathbf{j}$ appearing in response to a concentration gradient:
\begin{equation}\label{transdiff}
\mathbf{j} = - D_t  \nabla c.
\end{equation}
It is assumed that the concentration gradient is sufficiently small so linear response is valid. 
One can rewrite Eq.~\eqref{transdiff} in terms of the gradient of the chemical potential. The two gradients are related by $\nabla c = (\partial c / \partial \mu) \nabla \mu$. Since $c = \nav / V$ with $V$ the volume of one cavity, and $d \nav / d \mu = \beta (\nvar)$, as follows from Eq.~\eqref{eq::defnav}, this yields:
\begin{align}\label{eq::Dtv2}
\mathbf{j} = - D_t \beta \frac{\nav}{V} \frac{\nvar}{\nav} \nabla \mu.
\end{align}
The Maxwell-Stefan (MS) diffusion coefficient $D_{\mathrm{ms}}$ is defined as \cite{JPCCkrishna2009}
\begin{equation}\label{eq::defmsdiff} 
\mathbf{j} = - D_{\mathrm{ms}}  \beta c \nabla \mu.
\end{equation}
From Eqs.~\eqref{eq::Dtv2} and \eqref{eq::defmsdiff} it follows that
\begin{equation}\label{eq::DtDMS}
D_t =  \Gamma D_{\mathrm{ms}},
\end{equation}
where we have defined the thermodynamic factor:
\begin{equation}\label{eq::thdfactor}
\Gamma(\mu) = \frac{\langle n \rangle}{\langle n^2 \rangle - \langle n \rangle^2}.
\end{equation}

From Eq.~\eqref{eq::DtDMS} one can see that the transport diffusion is the product of a thermodynamic term $\Gamma$ and a kinetic term $D_{\mathrm{ms}}$. Because thermodynamic effects are ``factored out'', or corrected for, in $D_{\mathrm{ms}}$, it is sometimes called the corrected diffusion.
The Maxwell-Stefan diffusion coefficient can be written as
\begin{equation}\label{eq::greenkubopos}
D_{\mathrm{ms}} = \lim_{t \rightarrow \infty} \frac{1}{2 d N t} \overline{ \left( \sum_{i=1}^N  \Delta \mathbf{r}_i(t)  \right)^2 },
\end{equation}
where the sum runs over all $N$ particles in the system.
This definition is similar to the one for the self-diffusion Eq.~\eqref{eq::selfdiff}, with the MSD of $N$ times the center of mass of all particles instead of the MSD of one particle.
$D_{\mathrm{ms}}$ is therefore also called the center-of-mass diffusion coefficient.
From Eqs.~\eqref{eq::DtDMS} and \eqref{eq::greenkubopos} one finds the following relation between the self- and transport diffusion:
\begin{align}
D_t &=   \lim_{t \rightarrow \infty} \frac{\Gamma}{2 d N t} \left(  \sum_i \overline{ \Delta \mathbf{r}^2_i  } + \sum_{i,j \neq i} \overline{ \Delta \mathbf{r}_i . \Delta \mathbf{r}_j } \right) \label{eq::DtDsline2} \\
&= \Gamma D_s  +  \lim_{t \rightarrow \infty} \frac{\Gamma}{2 d N t} \sum_{i,j \neq i} \overline{ \Delta \mathbf{r}_i(t) . \Delta \mathbf{r}_j(t) }. \label{eq::dsdtintpartcorr}
\end{align}
For conciseness we do not write the time dependence in Eq.~\eqref{eq::DtDsline2}.

Two types of correlation effects influence the diffusion. The first type considers only a \textit{single} particle. If the direction and average rate of subsequent jumps of a single particle are uncorrelated, the self-diffusion is equal to $D_s = \lambda^2 k_{\mathrm{av}} / 2 d$, with $k_{\mathrm{av}}$ the average jump rate. In general, however, subsequent jumps are correlated. Consider, for example, the case where only one particle can occupy each lattice site. If a particle jumps, it is more likely to return to the site from where it came, because this site is more likely to be empty. These single-particle correlations influence the self-diffusion $D_s$.
The second type considers the correlation between jumps of \textit{different} particles. It is described by the second term on the right-hand side of Eq.~\eqref{eq::dsdtintpartcorr}. If the particles have a tendency to drag along other particles, then this term is positive. This happens, for example, when there is interparticle friction. The Maxwell-Stefan theory of diffusion is often used to study diffusion in porous materials \cite{JPCCkrishna2009}. In this context one can derive the relation: \begin{equation}
\frac{1}{D_s} = \frac{1}{D_{\mathrm{ms}}} + \frac{1}{D_{\mathrm{cor}}}.
\end{equation}
Interparticle correlations are captured by the term $1 / D_{\mathrm{cor}}$, while single-particle correlations influence $D_s$. $1 / D_{\mathrm{cor}}$ is interpreted as resulting from interparticle friction (in continuum models) or correlations between jumps of different particles (in lattice models \cite{paschek2001}). It is positive if the interparticle correlation term in Eq.~\eqref{eq::dsdtintpartcorr} is positive and vice versa. 
\section{\label{sec::eqprop}Equilibrium Properties}
In this section we discuss the equilibrium properties of the model. Since the focus is here on the influence of the various interactions as compared to the ideal case, we write the free energy as
\begin{align}\label{eq::Fidealgas}
F(n) = F^{\mathrm{id}}(n) + f(n),
\end{align}
where $F^{\mathrm{id}}(n)$ is the free energy of an ideal gas:
\begin{equation}
F^{\mathrm{id}}(n) \equiv k T \left[ \ln (n!) - n \lnÊ\left( V / \Lambda^3  \right) \right],
\end{equation}
with $V$ the volume of a cavity and $\Lambda$ the thermal de Broglie wavelength. We call $f(n)$ the interaction free energy, which includes all interactions and confinement. The free energy can be derived from the partition function, defined by
\begin{align}
Z(n) &= \frac{V^n}{n! \Lambda^{3n}} z(n), \\
z(n) &= \frac{1}{V^n} \int_{V} d\mathbf{r}_1 \ldots  \int_{V}d\mathbf{r}_n e^{-Ê\beta U(\mathbf{r}_1, \ldots, \mathbf{r}_n)} \label{eq::confint},
\end{align}
with $\mathbf{r}_i$ the position of the $i^{\mathrm{th}}$ particle and $U(\mathbf{r}_1, \ldots, \mathbf{r}_n)$ the interaction energy. The interaction free energy is then determined by the configurational integral $z(n)$ through $f(n) = - kT \ln z(n)$.

Previous results \cite{PRLbecker} showed that particle clustering occurs if the interaction free energy is concave. The effect of $f(n)$ on fluctuations in particle number, and the equivalence of a concave $f(n)$ and clustering, is discussed in more detail in Sec.~\ref{sec::thdfactor}.

An investigation of particle clustering in porous materials using MD simulations can be found in Refs.~\cite{LANGMUIRkrishnaTc,LANGMUIRkrishna2010_1,LANGMUIRkrishnaHbond}. If particles cluster the inverse thermodynamic factor $\Gamma^{-1}$ is larger than 1, and there are steep adsorption isotherms. Porous materials are classified by the characteristic dimensions of their structure. If the pore dimensions are smaller than 2 nm, then the material is called microporous; if the dimensions are between 2 and 50 nm, then it is called mesoporous; and even larger pores are called macroporous. Steep isotherms and $\Gamma^{-1} > 1$ are more common in macro- and mesoporous materials than in microporous materials, i.e., the more confining the geometry the less likely particle clustering occurs. 
All these features are present in our model and can be understood from the shape of the interaction free energy, as discussed in this section.
\subsection{\label{sec::thdfactor}Fluctuations in particle number}

\begin{figure*}%
\centering
\subfloat[][]{\includegraphics[width=0.9\columnwidth]{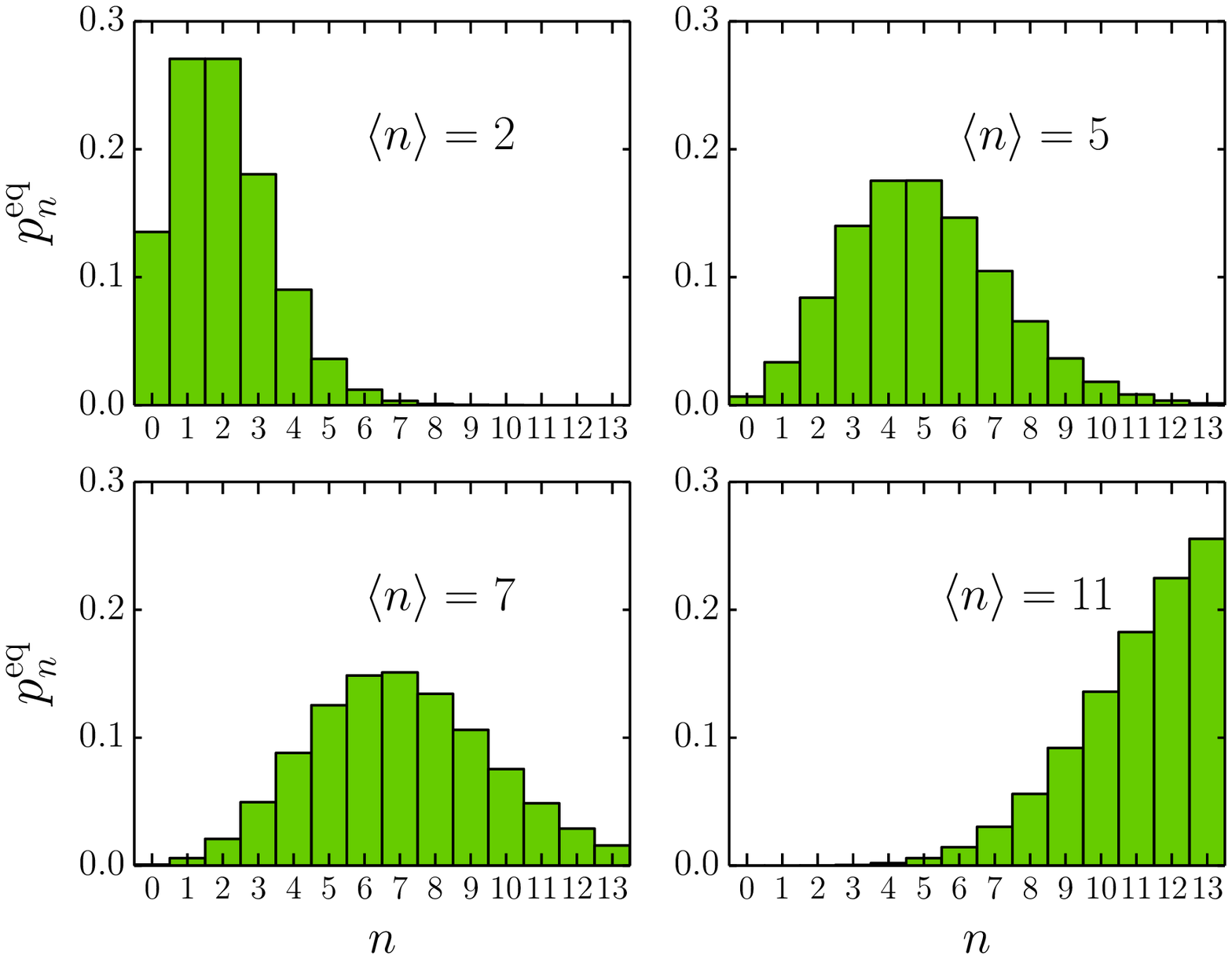}
\label{fig::histogram_fn0}
 }
\subfloat[][]
{
\includegraphics[width=0.9\columnwidth]{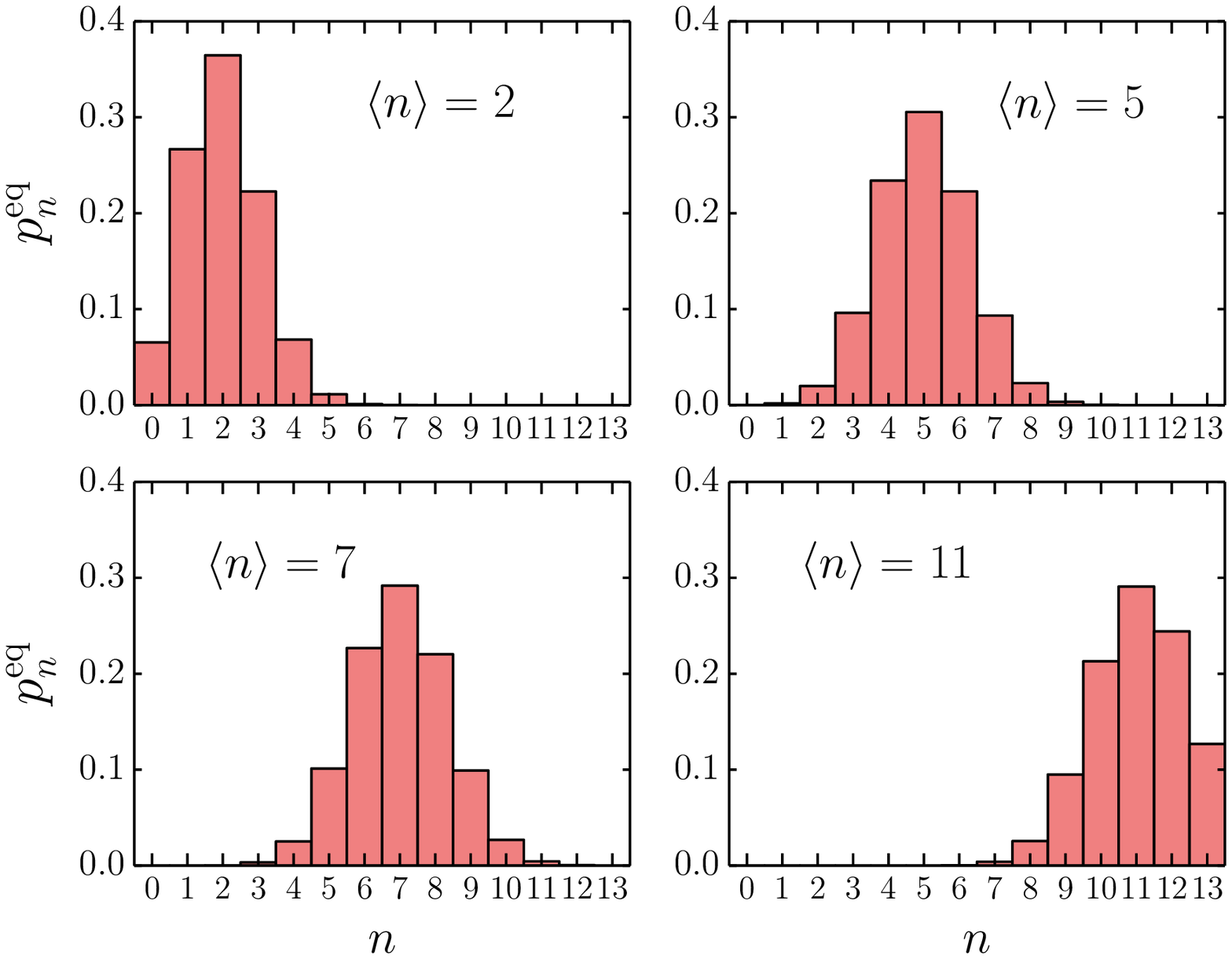}
\label{fig::histogram_fnconvex}
}
\\
\subfloat[][]
{
\includegraphics[width=0.9\columnwidth]{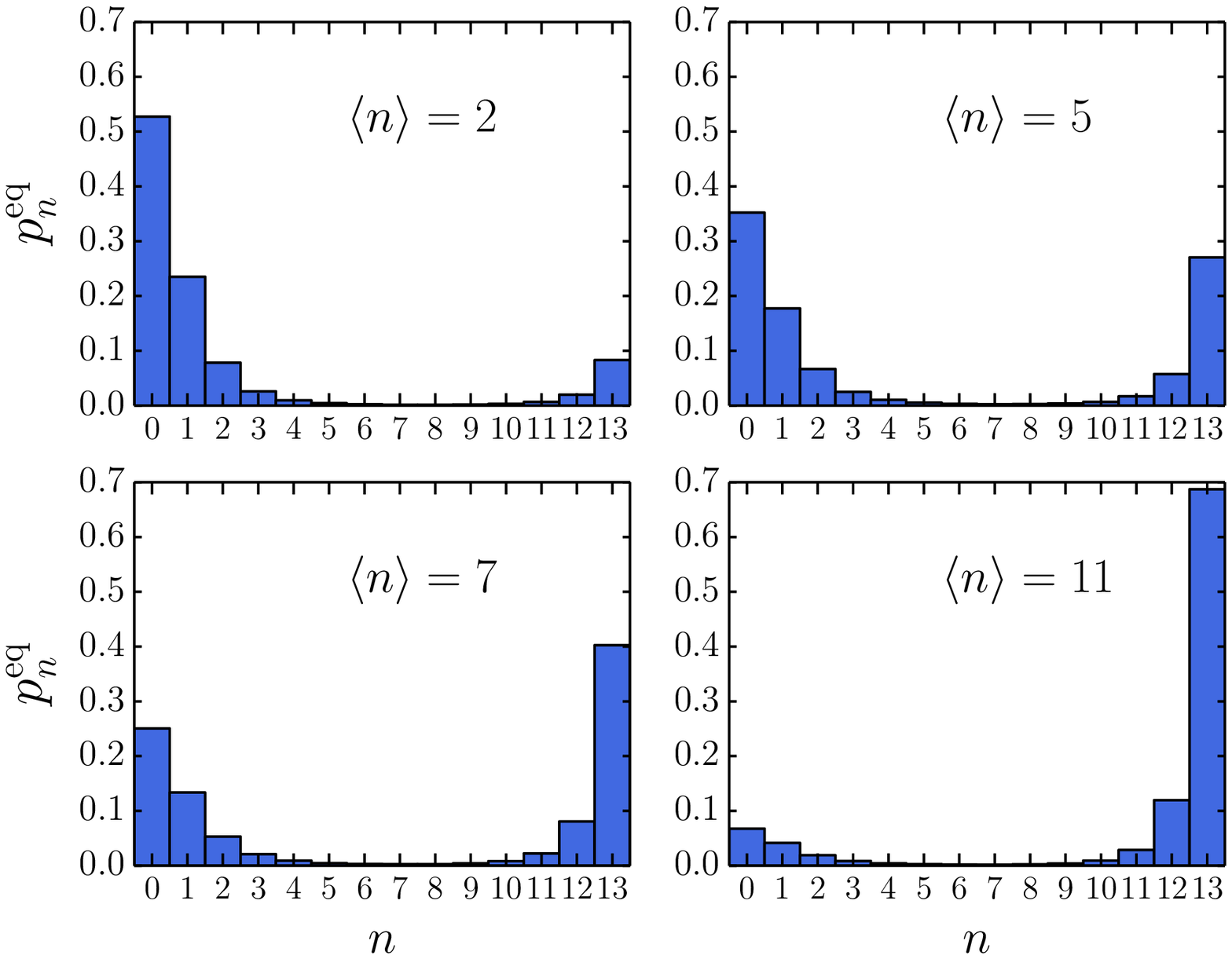}
\label{fig::histogram_fnconcave}
 }
\subfloat[][]
{
\includegraphics[width=0.9\columnwidth]{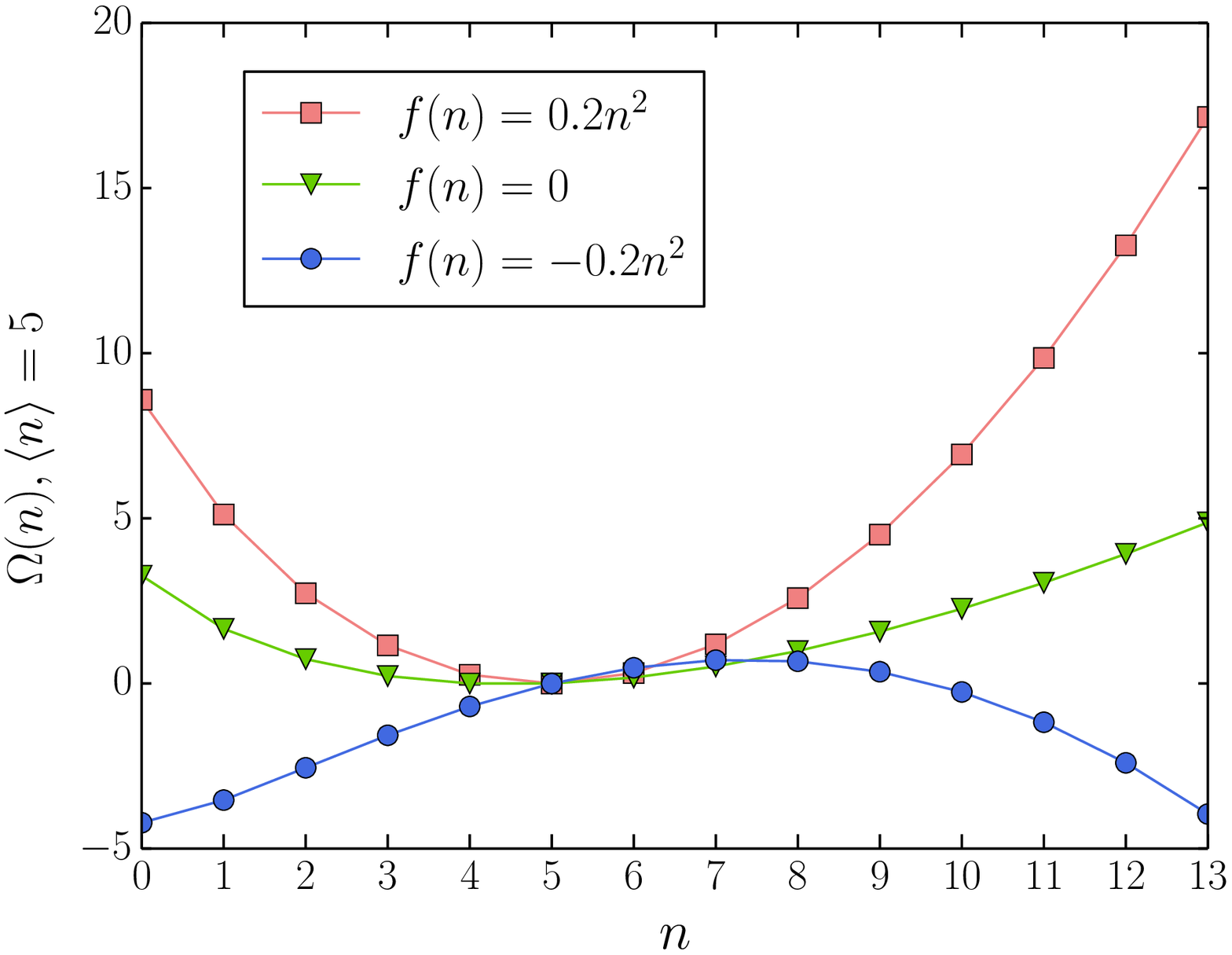}
\label{fig::Gn_nav5}
}
\caption{(Color online) The probability distribution of cavity occupation $p^{\mathrm{eq}}_n$, Eq.~\eqref{eq::probeq}, at different loadings and with $\nmax = 13$, for $(a)$ $f(n) = 0$ $(b)$ $f(n) = 0.2 n^2$ $(c)$ $f(n) = -0.2 n^2$. $(d)$ $\Omega(n)$ for different interactions, at $\langle n \rangle = 5$. The lines are a guide to the eye.}%
\label{fig:histo_and_Gn}%
\end{figure*}

\begin{figure}
\centering
\includegraphics[width=0.9\columnwidth]{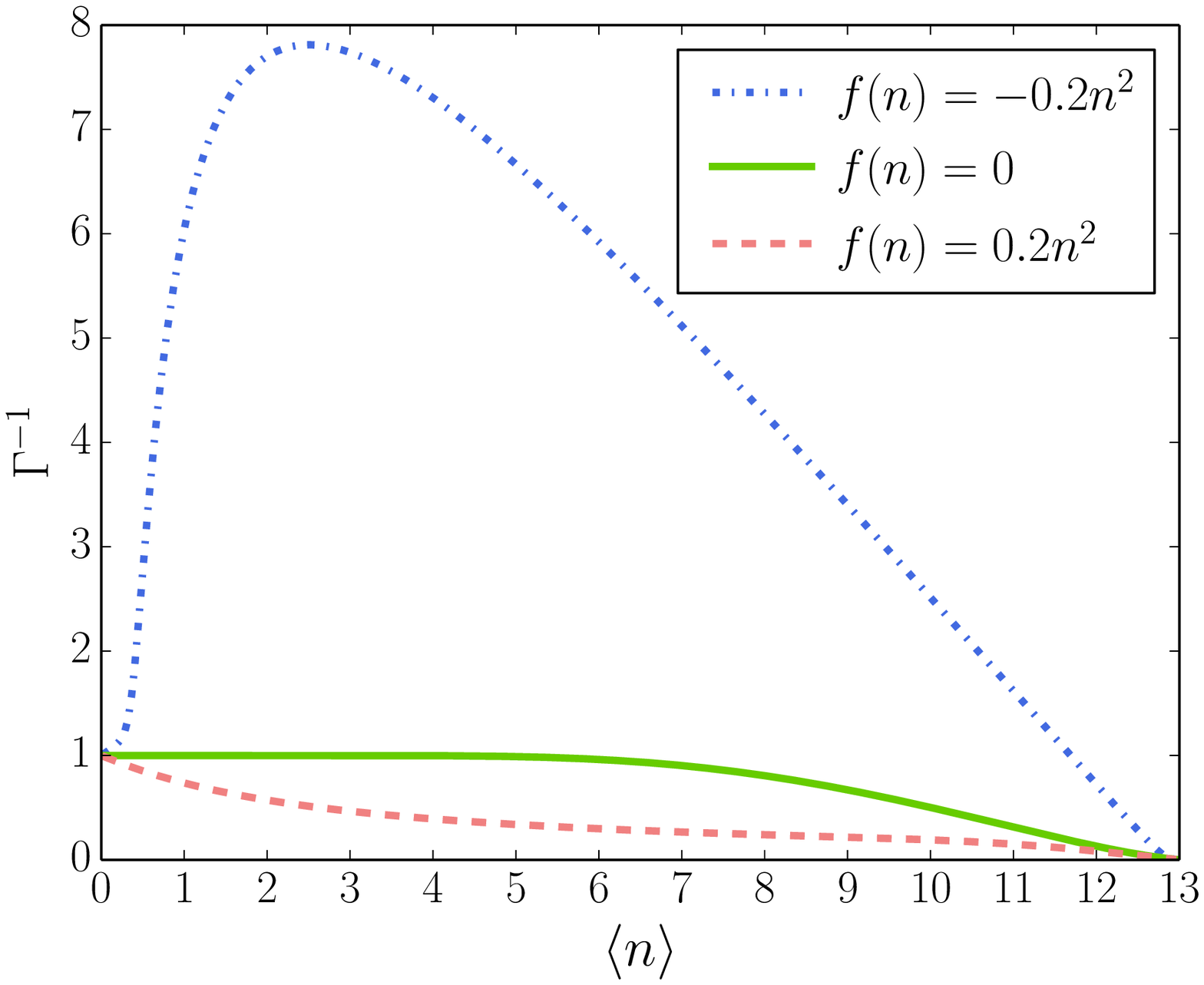}
\caption{(Color online) The inverse thermodynamic factor $\Gamma^{-1} = \left( \langle n^2 \rangle - \langle n \rangle^2 \right) / \nav$, for $\nmax = 13$ and different interactions.}
\label{fig::TF_interaction}
\end{figure}

Fluctuations in particle number are encoded in the thermodynamic factor, Eq.~\eqref{eq::thdfactor}.
We discuss the behavior of $\Gamma^{-1}$ instead of $\Gamma$ because the latter goes to infinity at maximum loading, making it more difficult to interpret graphically. 
Let us first consider noninteracting particles with zero volume, i.e., $f(n)=0$ and $\nmax = \infty$.
The distribution $p_n^{\mathrm{eq}}(\mu)$ is then a Poisson distribution,
\begin{equation}\label{eq::poissondistr}
p^{\mathrm{eq}}_n(\mu) = \frac{\langle n \rangle^n}{n!} e^{- \langle n \rangle},
\end{equation}
for which $\langle n^2 \rangle - \langle n \rangle^2 = \langle n \rangle$ and $\Gamma = 1$ at all loadings.
As argued previously \cite{PRLbecker}, the change in variance $\langle n^2 \rangle - \langle n \rangle^2$ is caused by the convexity versus concavity of $f(n)$. 
For better insight into the dependence of the equilibrium distributions on this convexity or concavity of $f(n)$ we consider the grand potential $\Omega(n,\mu)$, which we write as
\begin{align}
\Omega(n,\mu) = F^{\mathrm{id}}(n) + f(n) - \mu n, \label{eq::gnifvfn}
\end{align}
or, by defining the constant $\mu_0 \equiv k T \lnÊ\left( V / \Lambda^3  \right)$,
\begin{equation}
\Omega(n,\mu) = k T \ln \left( n! \right) + f(n) - \left( \mu + \mu_0 \right) n.
\end{equation} 
$\mu_0$ changes the chemical potential for which a certain loading is achieved; it does not change the behavior of $\Omega(n, \mu)$ at a given loading. Because we only consider isothermal systems $kT$ is a constant, which we take to be equal to 1.
We discuss three situations: no interactions $f(n) = 0$; convex interactions, with as an example $f(n) = 0.2 n^2$; and concave interactions, with as an example $f(n) = - 0.2 n^2$. For all interactions we take $\nmax = 13$, corresponding to $f(n) = \infty$ for $n > \nmax$.

To understand the effect of introducing an $\nmax$, we consider the situation $f(n) = 0$ and $\nmax = 13$. The probability distributions of cavity occupation $p^{\mathrm{eq}}_n$ at different loadings are plotted in Fig.~\ref{fig::histogram_fn0}. As long as $p^{\mathrm{eq}}_{\nmax} \approx 0$, $p^{\mathrm{eq}}_n$ is equal to the Poisson distribution Eq.~\eqref{eq::poissondistr} and $\Gamma = 1$. Once the probability to be full becomes nonzero the variance decreases compared to the Poisson distribution, resulting in $\Gamma^{-1} < 1$; see Fig.~\ref{fig::TF_interaction}. 

To understand the change in variance for different interactions we plot $\Omega(n,\mu)$ as a function of $n$, at the chemical potentials for which $\langle n \rangle = 5$; see Fig.~\ref{fig::Gn_nav5}. For graphical clarity the three curves are shifted vertically so $\Omega(5) = 0$. 
For $f(n) = 0$ the minimum of $\Omega(n)$ lies at $n = \nav$. All other values of $\Omega(n)$ are higher, because $\Omega(n) = k T \ln \left( n! \right) - (\mu + \mu_0) n$ is a convex function of $n$.
By adding a convex $f(n)$, $\Omega(n)$ increases faster around its minimum, and therefore all states that differ from $n = \nav$ become less likely compared to $f(n) = 0$. This is clear from the probability distributions of cavity occupation for $f(n) = 0.2 n^2$; see Fig.~\ref{fig::histogram_fnconvex}. As a result, $\Gamma^{-1} < 1$ at all loadings, cf.~Fig.~\ref{fig::TF_interaction}. An example of a convex $f(n)$, and hence a concave $z(n)$, is discussed in Ref.~\cite{CESTunca2003}, where it was attributed to excluded volume interactions between the methane molecules.

Adding a concave $f(n)$ gives the opposite behavior. $\Omega(n)$ increases more slowly around the average, and occupations that differ from the average become more likely compared to $f(n)=0$. For a very concave $f(n)$, $\Omega(n)$ no longer has a single minimum around $n = \nav$; there are two minima, at $n=0$ and $n=\nmax$; see Fig.~\ref{fig::Gn_nav5}. The probability distributions of cavity occupation for $f(n) = - 0.2 n^2$ are shown in Fig.~\ref{fig::histogram_fnconcave}. The particles cluster: The cavities are mostly empty or full. As a result, $\Gamma^{-1} > 1$ for low and medium loadings, after which the effect of $\nmax$ becomes dominant; cf.~Fig~\ref{fig::TF_interaction}.
An example of a concave contribution to $f(n)$ is the energy of a cluster of particles feeling short-range attractive interactions, which scales as $\propto n^{2/3}$ for large $n$ \cite{LANGMUIRmossa2004}. An example of clustering is found for particles undergoing hydrogen bonding \cite{LANGMUIRkrishnaHbond}.

\subsection{\label{sec::adsiso}Adsorption isotherm}

\begin{figure*}
\centering
\includegraphics[width=1.5\columnwidth]{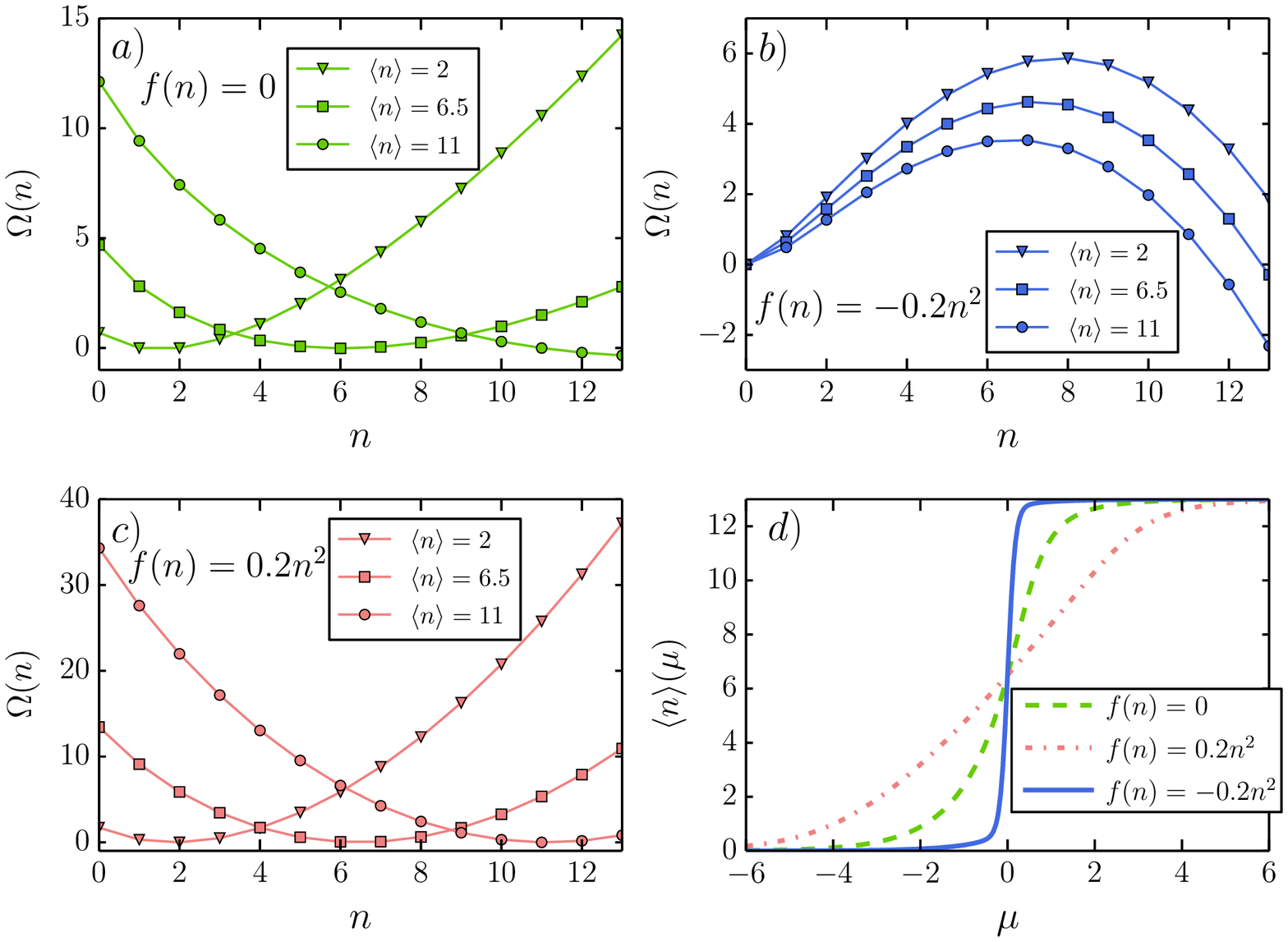}
\caption{(color online). $a), b), c)$ $\Omega (n)$ for different interactions and loadings, with $\nmax = 13$. For $f(n) = 0$ and $f(n) = 0.2 n^2$ the curves are shifted vertically so that $\Omega (\langle n \rangle) = 0$. For $f(n) = -0.2 n^2$ the curves are shifted vertically so that $\Omega (0) = 0$. The lines are a guide to the eye. $d)$ Adsorption isotherms $\langle n \rangle (\mu)$. The curves are shifted horizontally so that $\langle n \rangle (0) = 0.5 \nmax$.}
\label{fig::adsiso_gn}
\end{figure*}

\begin{figure}
\centering
\includegraphics[width=0.9\columnwidth]{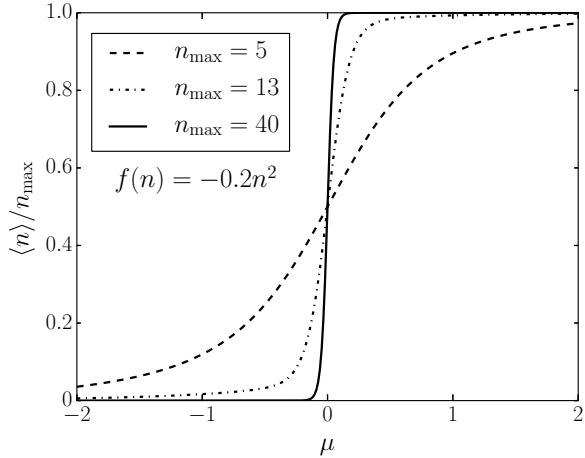}
\caption{Adsorption isotherms $\langle n \rangle (\mu) / \nmax $ for different values of $\nmax$, for the interaction $f(n) = - 0.2 n^2$. The curves are shifted horizontally to have $\langle n \rangle (0) = 0.5 \nmax$.}
\label{fig::adsiso_nmax}
\end{figure}
\begin{figure}
\centering
\includegraphics[width=1.0\columnwidth]{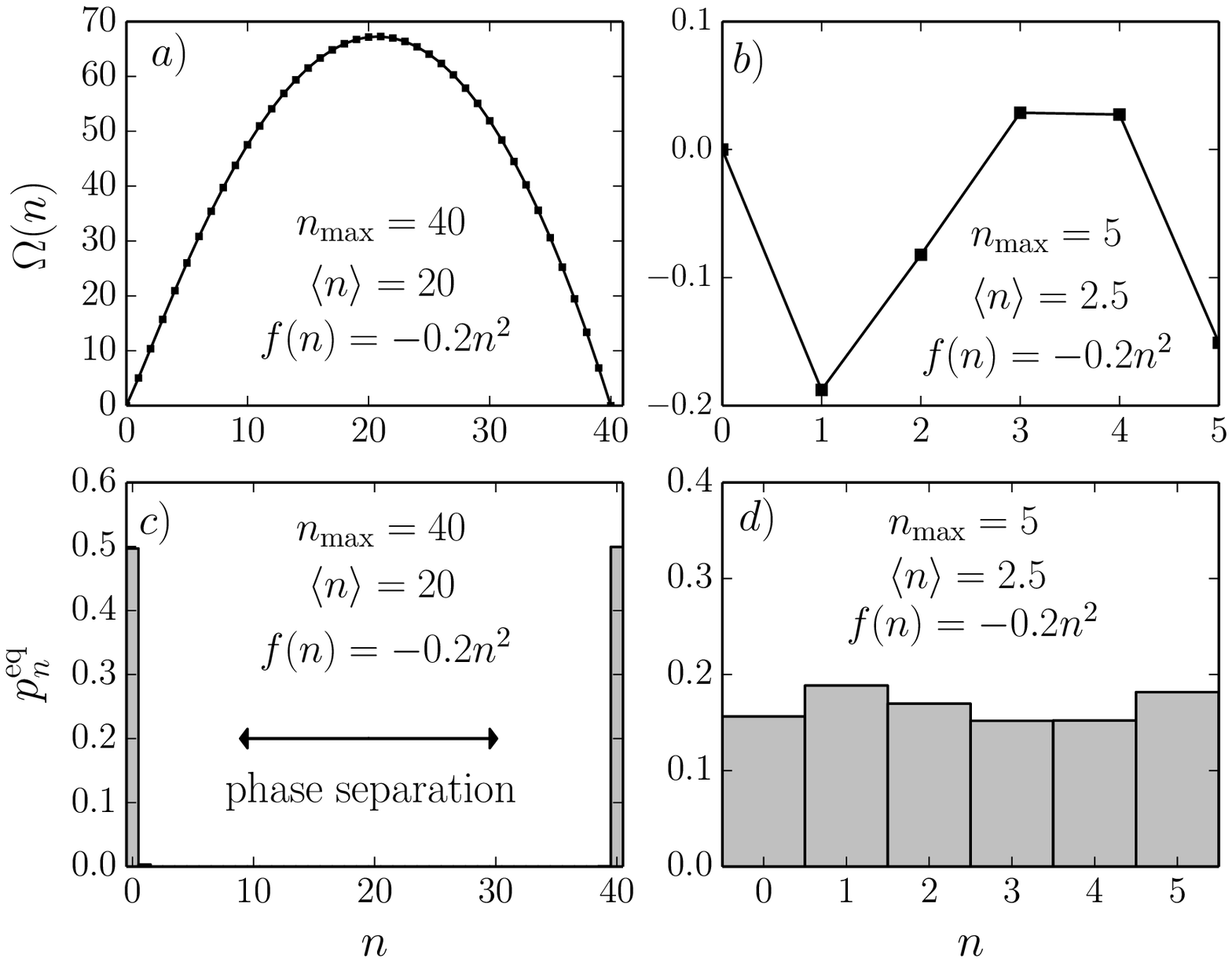}
\caption{$a)$ $\Omega(n)$ and $c)$ $p^{\mathrm{eq}}_n$, for $\nmax = 40$, $\nav = 20$, and $f(n) = - 0.2 n^2$. $b)$ $\Omega(n)$ and $d)$ $p^{\mathrm{eq}}_n$ for $\nmax = 5$, $\nav = 2.5$, and $f(n) = - 0.2 n^2$.}
\label{fig::Gn_histo_nmax5and40}
\end{figure}
\begin{figure}
\centering
\includegraphics[width=0.9\columnwidth]{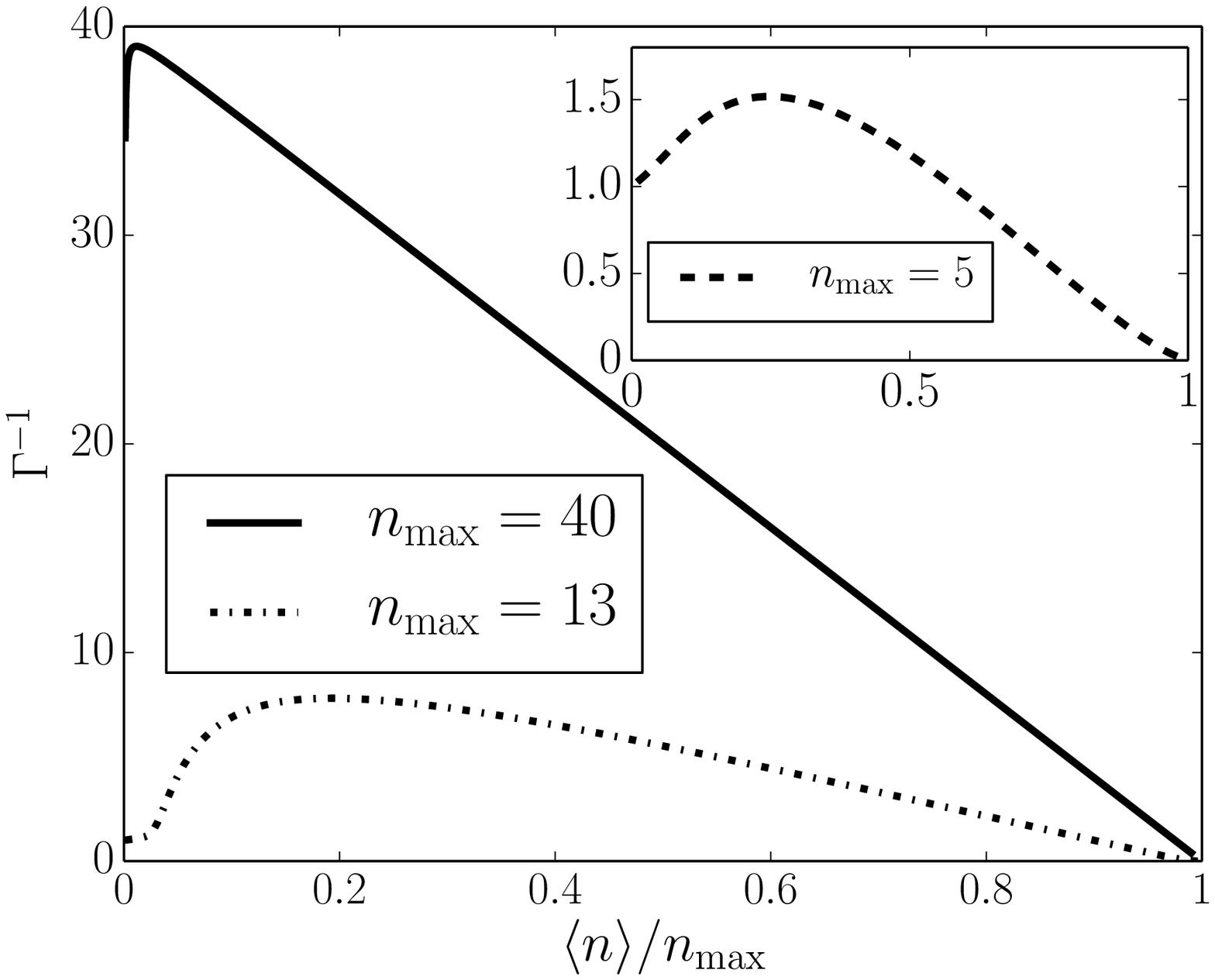}
\caption{The inverse thermodynamic factor $\Gamma^{-1} = \left( \langle n^2 \rangle - \langle n \rangle^2 \right) / \nav$, for different $\nmax$ and $f(n) = -0.2 n^2$.}
\label{fig::TF_nmax}
\end{figure}

Adsorption isotherms give the equilibrium concentration of particles in the system as function of, e.g., the pressure or chemical potential of the reservoir. 
Studies of adsorption in porous materials which use the same model assumptions as presented in this paper have been performed by several authors. In these studies one tries to predict and explain the behavior of the adsorption isotherm, using only a few microscopic parameters that describe the particle-particle and particle-cavity interaction. One of the first such analyses was performed by Ruthven \cite{NPSruthven1971}. Similar studies have been performed both analytically \cite{JCPayappa1,JCPayappa2,Langmuirbae,JCPdemontis2009,LANGMUIRgarcia2013,JPCCgarcia} and numerically \cite{CESTunca2003,JCPvantassel,Langmuirvantassel}. We refer to Ref.~\cite{BOOKruthven} for an introduction. We discuss here the qualitative influence of $f(n)$ and $\nmax$ on the adsorption isotherm.
The particle concentration is equal to $\langle n \rangle / \lambda^d$ (with $d$ the dimension). Since the term $\lambda^d$ only rescales the adsorption isotherm by a constant, we study $\langle n \rangle(\mu)$.

The adsorption isotherms $\langle n \rangle(\mu)$ for the three considered interactions are plotted in Fig.~\ref{fig::adsiso_gn}d. The grand potentials at different loadings are shown in Figs.~\ref{fig::adsiso_gn}a,b,c. The adsorption isotherm for the concave $f(n)$ is steeper than the one for $f(n) = 0$, which is steeper still than the one for the convex $f(n)$. Such behavior can be understood from $\langle n \rangle (\mu)$: 
\begin{equation}\label{eq::steepvsvar}
\frac{d \nav}{d \mu} = \beta \left( \nvar \right) = \beta \nav \Gamma^{-1}.
\end{equation}
A concave $f(n)$ leads to a larger value of $\Gamma^{-1}$, which means a steeper adsorption isotherm.
Steep isotherms occur if there is a first-order phase transition, for example, when there is capillary condensation \cite{RPPgelb1999}. They have also been found for systems where the particles cluster; see Ref.~\cite{LANGMUIRkrishna2010_1} and references therein. 
The connection among clustering, first-order phase transitions, and steep isotherms can be understood from Fig.~\ref{fig::adsiso_gn}. For noninteracting and repulsive particles (Figs.~\ref{fig::adsiso_gn}a,c) only the average concentration $\nav$ is stable (i.e., a local minimum). Increasing the chemical potential gradually shifts this local minimum to higher concentrations. For the concave $f(n)$, Fig.~\ref{fig::adsiso_gn}b, there are in contrast two stable concentrations, at $n=0$ and $n=\nmax$. Increasing the chemical potential causes a sudden shift of the global minimum from $n=0$ to $n=\nmax$, resulting in a steep isotherm.

While stepped isotherms are commonly encountered in mesoporous materials, they are quite rare in microporous materials \cite{LANGMUIRkrishna2010_1}. This can be seen as the result of the confinement, which prevents the formation of stable macroscopic phases \cite{RPPgelb1999,PRE_melnichenko,JML_melnichenko}. We discuss how such behavior is reproduced in our model.

We take the diameter of the cavities as the characteristic dimension of the system (i.e., $\nmax$). Consider the same type of particles in materials with cavities of different size. The smaller the volume of the cavities the lower $\nmax$. To study the transition from micro- to macroporous for clustering particles, we consider the interaction $f(n) = - 0.2 n^2$, for $\nmax = 5, 13,$ and $40$. The adsorption isotherms are presented in Fig.~\ref{fig::adsiso_nmax}. Their steepness decreases with decreasing cavity size. For an interpretation of this behavior we plot $p^{\mathrm{eq}}_n$ and $\Omega(n)$ at loading $\langle n \rangle / \nmax = 0.5$, for $\nmax = 40$ and $\nmax = 5$ in Fig.~\ref{fig::Gn_histo_nmax5and40}. For $\nmax = 40$ a stable cluster consists of 40 particles, and the thermodynamic barrier between the stable phases $n=0$ and $n=40$ is very large, cf.~Fig.~\ref{fig::Gn_histo_nmax5and40}a. Fluctuations between the two phases are highly unlikely, as can be seen from $p^{\mathrm{eq}}_n$ in Fig.~\ref{fig::Gn_histo_nmax5and40}c. This resembles the situation where there is a macroscopic phase separation (i.e., in the thermodynamic limit), where moving on the infinitely steep part of the adsorption isotherm corresponds to changing the relative portion of the two phases of the system. 
For $\nmax = 5$ the maximum size of a cluster is 5 particles. Such a cluster is easily broken by fluctuations. In fact, the grand potential does not show the typical structure of two stable minima [see Fig.~\ref{fig::Gn_histo_nmax5and40}b], and the adsorption isotherm shows no real steepness. 
$\nmax = 13$ is an intermediate case of these two situations; see Figs.~\ref{fig::histogram_fnconcave} and \ref{fig::adsiso_gn}b.
The inverse thermodynamic factors are plotted in Fig.~\ref{fig::TF_nmax}. 
For $\nmax = 40$, $\Gamma^{-1}(\nav)$ is approximately a straight line between $\Gamma^{-1}(0) \approx \nmax$ and $\Gamma^{-1}(\nmax) = 0$. This can be understood by making the approximation that cavities are either empty or full. In this case, $p_{\nmax}  = \nav / \nmax$, $p_{0}  = 1- \nav / \nmax$, and
\begin{equation}
\Gamma^{-1}(\nav) = \nmax \left( 1 - \frac{\nav}{\nmax} \right).
\end{equation}
If particle clustering occurs the inverse thermodynamic factor goes from showing almost no increase above 1 for microporous materials, to a straight line between $\nmax$ at $\nav = 0$ and $0$ at $\nav = \nmax$ for macroporous materials, cf.~Fig.~\ref{fig::TF_nmax}.

\section{\label{sec::dynamic}Dynamical Properties}
In this section we discuss the dynamical properties of the model. In Sec.~\ref{sec::transrates} we use transition-state theory to calculate possible forms for the rates $k_{nm}$. 
In Ref.~\cite{PRLbecker} we obtained analytical expressions for $D_s$ and $D_t$ for a system of length $L=1$. In Sec.~\ref{sec::DMFA} we show that the same expressions are obtained in an infinitely long system if one ignores all correlations. In Sec.~\ref{sec::diffandcorr} we investigate the diffusion properties of the model.
\subsection{\label{sec::transrates}Transition rates}

\begin{figure}
\centering
\includegraphics[width=0.6\columnwidth]{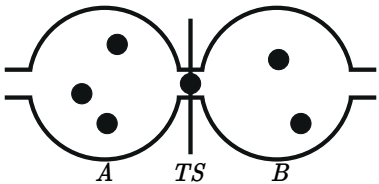}
\caption{Two cavities, A and B, divided by a transition-state surface TS. A particle is in the transition state.}
\label{fig::TST}
\end{figure}

The free energy $F(n)$ does not fully specify the dynamics, contained in the rates $k_{nm}$, because only local detailed balance Eq.~\eqref{eq::localdetbalance} has to be obeyed. For example, all rates of the form
\begin{equation}
k_{nm} = \nu n \frac{e^{- \beta c \left[ f(n-1) - f(n) \right]}}{e^{- \beta (1-c) \left[ f(m) - f(m+1) \right]}}, \label{eq::GDBrate}
\end{equation}
obey local detailed balance for any $c \in {\mathbb{R}}$ ($\nu$ denotes a positive constant throughout this paper). As discussed in Appendix \ref{app:tstcalc}, the physically relevant rates are found for $0 \leq c \leq 1$, where $c$ measures the importance of the interaction of the two cavities participating in the jump. We use TST \cite{RMPhanggi} to calculate possible forms of the jump rates. The details of the calculations can be found in Appendix \ref{app:tstcalc}.

Consider two connected cavities, called $A$ and $B$, containing respectively $n$ and $m$ particles. In the middle of the window we define a transition-state (TS) surface. If the center of a particle is located on the TS surface, it is said to be in the transition state. The setup is illustrated in Fig.~\ref{fig::TST}. The jump rate $k_{nm}$ is equal to the probability that a particle from cavity $A$ is in the transition state, multiplied by its average velocity towards cavity $B$.

Consider, first, particles which have no long-range interactions. An example is when the particles only feel hard-core repulsion. As a result, the particle in the TS has no influence on the interaction of the particles in cavities $A$ and $B$. The rates then have the form:
\begin{align}
k_{nm} = k_{10} e^{- \beta [ f(1)-f(0) ]} n e^{\betaÊ[ f(n) - f(n-1) ]}. \label{eq::rates1_2}
\end{align}
As always we require that $k_{n,\nmax} = 0$ for all $n$. This jump rate only depends on the change in free energy of cavity $A$. Note that it corresponds to $c = 1$ for the rate given in Eq.~\eqref{eq::GDBrate}.

Consider now particles with long range interactions. The particle in the TS interacts with the particles of both cavities $A$ and $B$. $k_{nm}$ therefore depends on the change in interaction free energy of both cavities $A$ and $B$. We study rates of the form (see Appendix \ref{app:tstcalc}):
\begin{align}\label{eq::rates2}
k_{nm} = k_{10} n e^{- ( \beta / 2 ) \left[ f(n-1) + f(m+1) - f(n) - f(m) \right]}.
\end{align}
These are the rates used in Ref.~\cite{PRLbecker}, with $k_{10} = \nu$.  Note that the change in free energy of cavities $A$ and $B$ is of equal importance. This rate corresponds to $c = 1/2$ in Eq.~\eqref{eq::GDBrate}.

\subsection{\label{sec::DMFA}Dynamical mean-field approximation}

Consider an infinitely large equilibrium system at chemical potential $\mu$. The lattice is cubic and has dimension $d$, i.e., each cavity has $2d$ neighbors. We tag one particle at time $t=0$, and calculate its average MSD in the limit $t \uparrow \infty$. 
Subsequent jumps of the tagged particle are correlated because of memory effects in the environment (i.e., the other particles), as already explained in Sec.~\ref{sec::diffth}. For example, for $\nmax = 1$, the tagged particle is more likely to jump back to its previous position, because this cavity is more likely to be empty. The influence of such memory effects is discussed in detail in Sec.~\ref{sec::diffandcorr}.
In the dynamical mean-field (DMF) approximation all memory effects, or correlations between particle jumps, are neglected \cite{ADVPHYSalanissila2002}.
This assumption is equivalent to assuming that, after a jump of the tagged particle, the environment loses its memory instantly. Because the environment is memoryless, the cavities connected to the cavity containing the tagged particle have the equilibrium distribution $\peq$ at all times. We calculate $\hat{p}_n$, the probability that the cavity containing the tagged particle has $n$ particles in total (including the tagged particle). The tagged particle jumps away from a cavity containing $n$ particles to a cavity containing $m$ particles with rate $k_{nm}/n$. In the DMF approximation, the master equation for $\hat{p}_n$ reads:
\begin{align}
\dot{\hat{p}}_n = &\sum_{m=1}^{\nmax} \hat{p}_m p^{\mathrm{eq}}_{n-1} 2 d \frac{k_{m,n-1}}{m}  - \sum_{m=0}^{\nmax-1} \hat{p}_n p^{\mathrm{eq}}_m 2 d \frac{k_{nm}}{n} \notag \\
+ &\sum_{m=1}^{\nmax} \hat{p}_{n-1} p^{\mathrm{eq}}_{m} 2 d k_{m,n-1} - \sum_{m=0}^{\nmax-1} \hat{p}_n p^{\mathrm{eq}}_m 2 d k_{nm}\frac{n-1}{n} \notag \\
+ &\sum_{m=0}^{\nmax-1} \hat{p}_{n+1} p^{\mathrm{eq}}_m 2d k_{n+1,m} \frac{n}{n+1} - \sum_{m=1}^{\nmax} \hat{p}_n p^{\mathrm{eq}}_m  2 d k_{mn}.
\end{align}
The positive terms are transitions toward the state $\hat{p}_n$: a jump of the tagged particle to a cavity containing $n-1$ particles (first line), a particle jump to the cavity containing the tagged particle from the state $\hat{p}_{n-1}$ (second line), and a particle jump away from the cavity containing the tagged particle from the state $\hat{p}_{n+1}$ (third line). The negative terms are transitions away from $\hat{p}_n$: a jump of the tagged particle to another cavity (first line), a particle jump away from the cavity containing the tagged particle (second line), and a particle jump to the cavity containing the tagged particle (third line). 
The stationary solution is
\begin{equation}\label{eq::soltag}
\hat{p}_n = \frac{n}{\nav} \peq.
\end{equation}
This can be checked by filling in Eq.~\eqref{eq::soltag} in the master equation and realizing that the two terms on each line cancel each other because of local detailed balance Eq.~\eqref{eq::localdetbalance}.
The average jump rate $\hat{k}_{\mathrm{av}}$ of the tagged particle is equal to:
\begin{align}
\hat{k}_{\mathrm{av}} &= 2 d \sum_{n=1}^{\nmax} \sum_{m=0}^{\nmax-1} \frac{ k_{nm}}{n} \hat{p}_n p^{\mathrm{eq}}_m \\
&= 2 d \sum_{n=1}^{\nmax} \sum_{m=0}^{\nmax-1} k_{nm} \frac{\peq}{\nav} p^{\mathrm{eq}}_m  = 2 d \frac{\langle k \rangle}{\nav}.
\end{align}
The particle is performing a random walk on a $d$-dimensional lattice with average jump rate $\hat{k}_{\mathrm{av}}$. The self-diffusion is in this case equal to:
\begin{align}
D_s = \frac{\lambda^2}{2 d} \hat{k}_{\mathrm{av}} = \frac{\lambda^2 \langle k \rangle}{\langle n \rangle} \label{eq::dsnocorr}.
\end{align}
Because all particle jumps are assumed to be uncorrelated, the inter-particle correlation term in Eq.~\eqref{eq::dsdtintpartcorr} is zero and one finds:
\begin{equation}\label{eq::dtnocorr} 
D_t = \frac{\lambda^2 \langle k \rangle}{\langle n^2 \rangle - \langle n \rangle^2}.
\end{equation}
As a result, $D_{\mathrm{ms}} = D_s$ in the DMF approximation.
The self-diffusion Eq.~\eqref{eq::dsnocorr} and transport diffusion Eq.~\eqref{eq::dtnocorr} obtained from the DMF approximation are the same as calculated for a system of length $L = 1$ \cite{PRLbecker}.

If there is no $\nmax$ and if the jump rates only depend on the number of particles in the cavity from which the particle jumps ($k_{nm} = k_n$), the model is a zero-range process (ZRP) \cite{ZRPevansreview}. The rates then have the form \cite{EPJSTbecker}:
\begin{equation}\label{eq::ZRPgeneral}
k_{n} = \nu n e^{\beta [ f(n) - f(n-1)Ê]}.
\end{equation}
Systems with the rates of Eq.~\eqref{eq::rates1_2} that have no $\nmax$ are therefore ZRPs.  
If a stationary solution exists the DMF results Eqs.~\eqref{eq::dsnocorr} and \eqref{eq::dtnocorr} are correct for all lengths and all interactions \cite{EPJSTbecker}. A stationary solution does not exist if there is particle condensation, i.e., if the number of particles in the cavities grows indefinitely in time. This could, e.g., occur if there is a concave interaction free energy.

\subsection{\label{sec::numsim}Numerical Simulations}

\begin{figure}%
\centering
\subfloat[][]{\includegraphics[width=0.9\columnwidth]{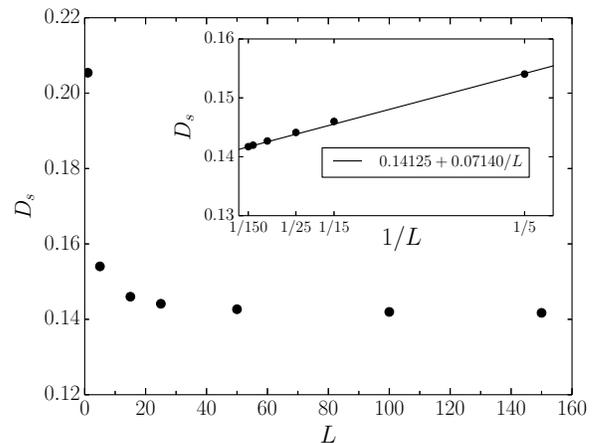}
\label{fig::Ds_L}
 }
\\
\subfloat[][]
{
\includegraphics[width=0.9\columnwidth]{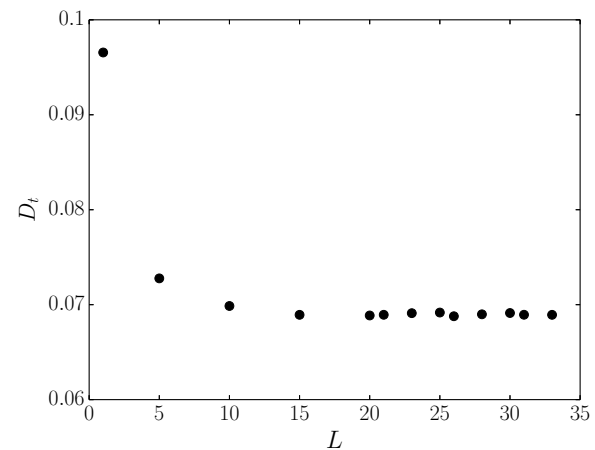}
\label{fig::Dt_L}
}
\caption{One-dimensional system with $f(n) = -0.2 n^2$, $\nmax = 13$, $\nav / \nmax = 0.8$, and the rates of Eq.~\eqref{eq::rates2}. $(a)$ Self-diffusion as a function of the length $L$ of the system. The inset shows the same data, plotted as a function of $1/L$. The analytical fit is obtained using Mathematica, and was done for all lengths except $L=1$. $(b)$ Transport diffusion.}%
\label{fig:DtandDs_L}%
\end{figure}

We discuss how the self- and transport diffusion are numerically simulated. The Markov dynamics of the system is simulated using the kinetic Monte Carlo method. The chemical potential corresponding to a certain loading $\nav$ can be found numerically via Eq.~\eqref{eq::defnav}. This chemical potential determines the rates at which particles are injected or removed at the boundaries; see Eq.~\eqref{eq::boundaryrates}.

To measure the self-diffusion at loading $\nav(\mu)$ both reservoirs are put at the chemical potential $\mu$. A concentration gradient of labeled particles is introduced by labeling particles that enter from the left or right reservoir with different percentages. $D_s$ can then be found using Eq.~\eqref{eq::selfdiff2}. In the simulations, all particles coming from the left reservoir are labeled ($100 \%$), and none of the particles coming from the right reservoir are labeled ($0 \%$). Taking different percentages gives the same $D_s$. We are interested in the situation where the boundary cavities have negligible influence.
The length dependence of $D_s$ in a one-dimensional system, for the parameters $f(n) = - 0.2 n^2$, $\nmax = 13$, the rates of Eq.~\eqref{eq::rates2}, at loading $\nav / \nmax  = 0.8$, is shown in Fig.~\ref{fig::Ds_L}. Once $L > 1$ the diffusion is influenced by correlations, and one observes a sharp decrease of $D_s$. The influence of the boundary cavities decreases with increasing length. For large $L$ the length dependence scales as $\propto 1/L$. This $1/L$ dependence can be increasing, decreasing, or (approximately) constant, depending on the loading of the system (data not shown).

The transport diffusion at loading $\nav$ is measured by putting the left and right reservoirs at different chemical potentials corresponding to, respectively, $\nav + \delta \nav$ and $\nav - \delta \nav$, where $\delta \nav$ should be small to ensure that one is in the regime of linear response. By measuring the particle flux through the system one can calculate $D_t$ using Eq.~\eqref{transdiff}.
The length dependence of the transport diffusion for the same parameters as $D_s$ is plotted in Fig.~\ref{fig::Dt_L}. It also shows a sharp decrease for small $L$ and remains almost constant after $L=15$. The time needed to achieve good statistics is much larger for the transport diffusion than for the self-diffusion. This is because the transport diffusion is measured for a small concentration gradient. For large $L$ the particle flux becomes very small, and the error bars on the transport diffusion are very large.
This is in contrast to $D_s$, where a concentration gradient of labeled particles is applied, which can be made arbitrarily high ($100 \%$ in the left reservoir and $0\%$ in the right reservoir). The problem of a very small labeled particle flux occurs much later compared to the transport diffusion. We have therefore not studied the length dependence of $D_t$ for large lengths.

\subsection{\label{sec::diffandcorr}Diffusion and correlations}
We discuss the diffusion for the two rates, Eqs.~\eqref{eq::rates1_2} and \eqref{eq::rates2}, for different interactions. We take $k_{10} e^{-\beta \left[ f(1) - f(0) \right]} = 1$ for the rates of Eq.~\eqref{eq::rates1_2} and $k_{10} = 1$ for the rates of Eq.~\eqref{eq::rates2}. In all cases $kT= \lambda = 1$. The other parameter values, and an explanation of how the error bars are determined, can be found in Appendix \ref{app:compdet}. We first discuss one-dimensional systems. Diffusion in two- and three-dimensional systems is considered at the end of the section.

Memory effects are studied by measuring directional correlations of subsequent jumps of a single particle, similarly to, e.g., Ref.~\cite{PRBvattulainen1999}. We tag one particle and record the direction of its first jump. We measure the probability $\tilde{p}(m)$ that its $m^{\mathrm{th}}$ jump has the same direction as its first jump. If the jumps are uncorrelated one has for a one-dimensional system that $\tilde{p}(m) = 1/2$ for all $m \geq 1$. We also measure $p(n|\hat{n})$, the probability that a neighboring cavity has $n$ particles given there are $\hat{n}$ particles in the cavity of the tagged particle. In the DMF approximation one has that $p(n | \hat{n}) = p^{\mathrm{eq}}_n$. When we say that correlations increase or decrease the diffusion this is always with reference to the DMF situation. The self-diffusion is influenced by correlations of subsequent jumps of a single particle. By comparing the ratio of the self- and transport diffusion with $\Gamma$ we have access to interparticle correlations, cf.~Eq.~\eqref{eq::dsdtintpartcorr}.

For $\nmax = 1$ our model reduces to the well-known Langmuir gas model \cite{taylor1933}. In an infinitely long one-dimensional equilibrium system single-file diffusion occurs \cite{kwinten2012epl,kwinten2012pre85}, resulting in $\langle x(t) \rangle \propto \sqrt{t}$ and $D_s = 0$. In higher-dimensional systems the diffusion is normal. The self-diffusion is lowered because of the \textit{back-correlation mechanism}: If a particle jumps, the cavity it came from is more likely to be empty, making it more likely that the particle jumps back. The transport diffusion is equal to the DMF value for all loadings: $D_t = k_{10} = 1$ \cite{RPPgomer1990}.

\begin{figure*}%
\centering
\subfloat[][]{\includegraphics[width=1\columnwidth]{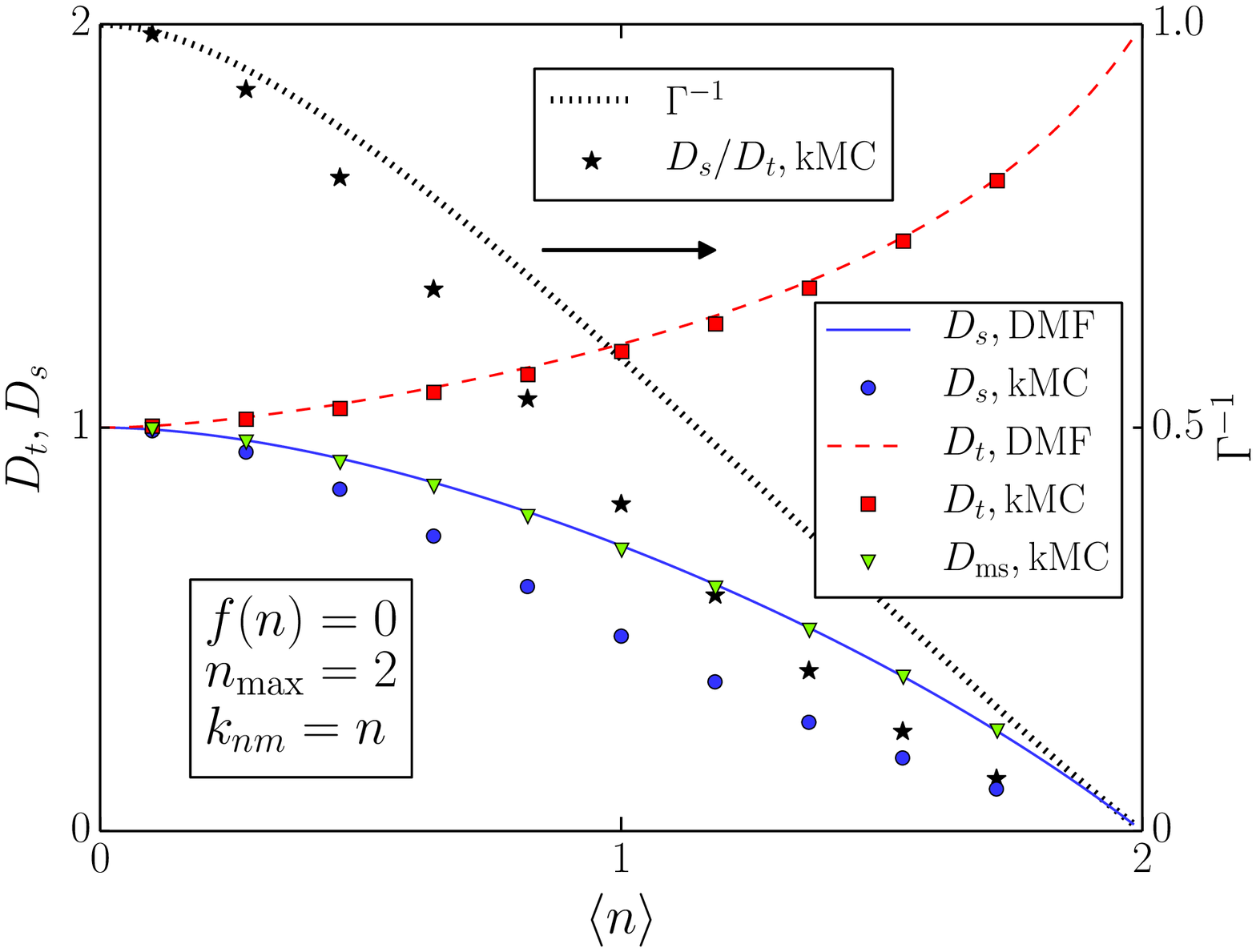}
\label{fig::diff_fn0_nmax2}
 }
\subfloat[][]
{
\includegraphics[width=1\columnwidth]{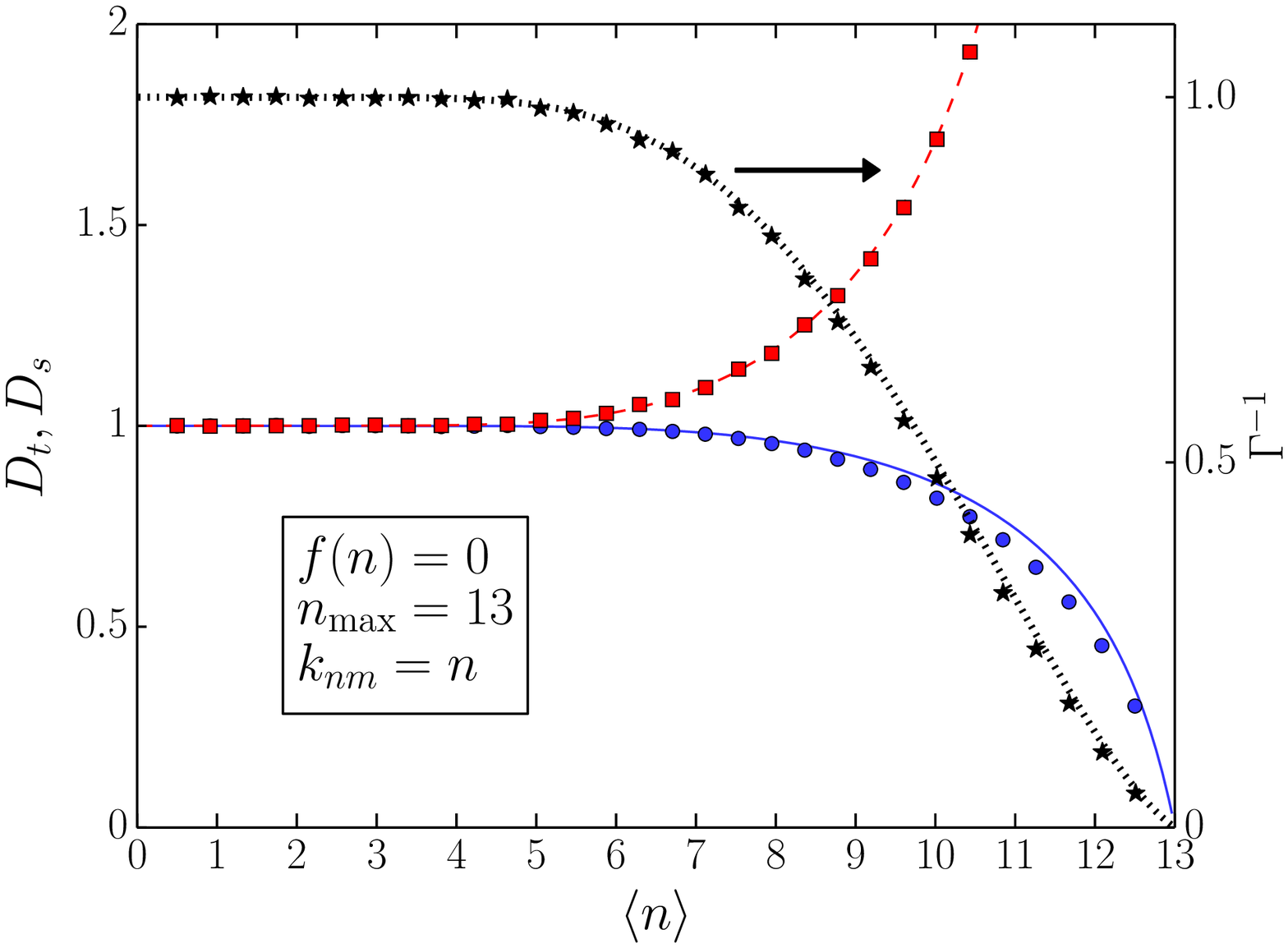}
\label{fig::diff_fn0}
}
\\
\subfloat[][]{\includegraphics[width=1\columnwidth]{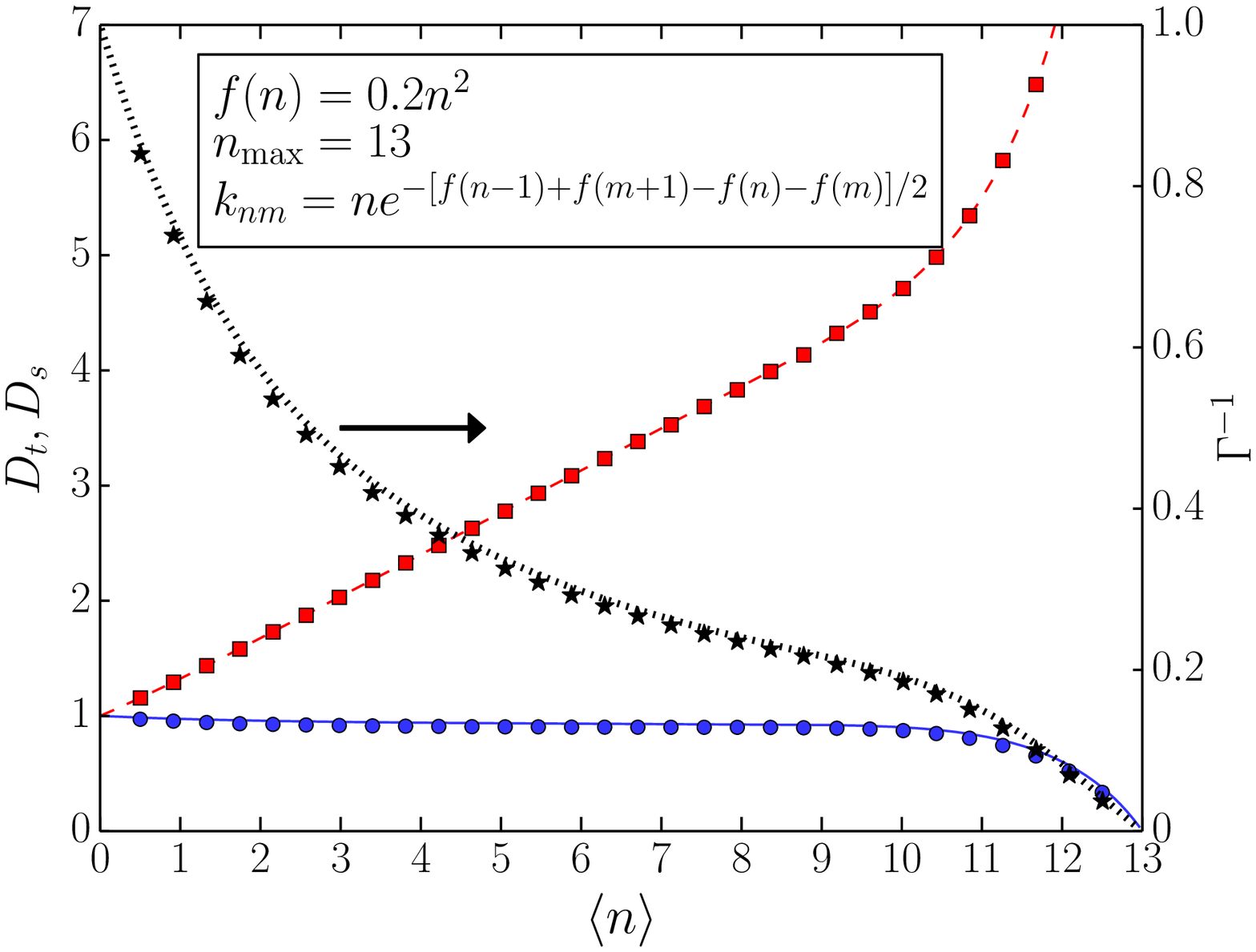}
\label{fig::diff_fnconvex_r2}
 }
\subfloat[][]
{
\includegraphics[width=1\columnwidth]{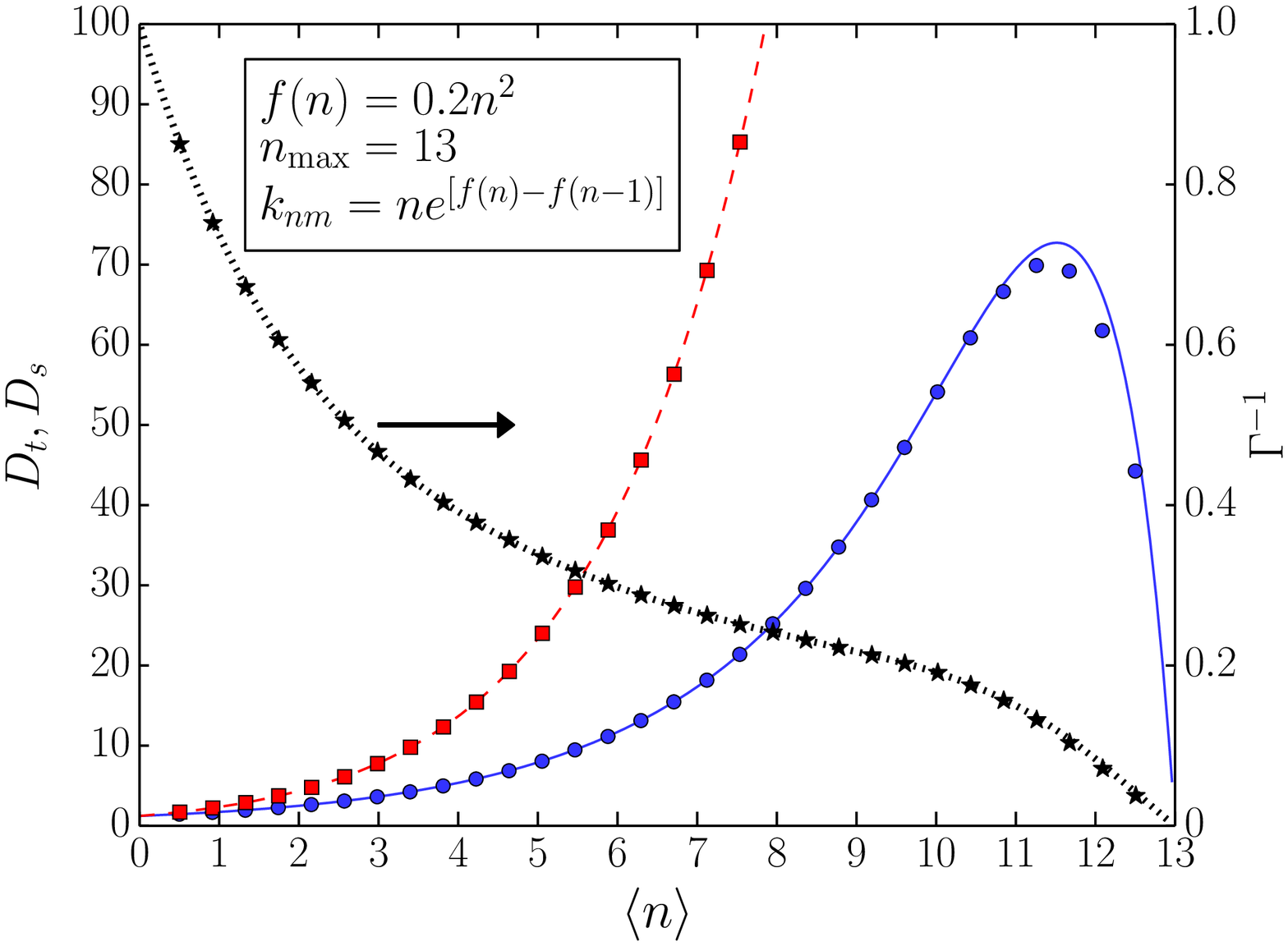}
\label{fig::diff_fnconvex_r1}
}
\\
\subfloat[][]{\includegraphics[width=1\columnwidth]{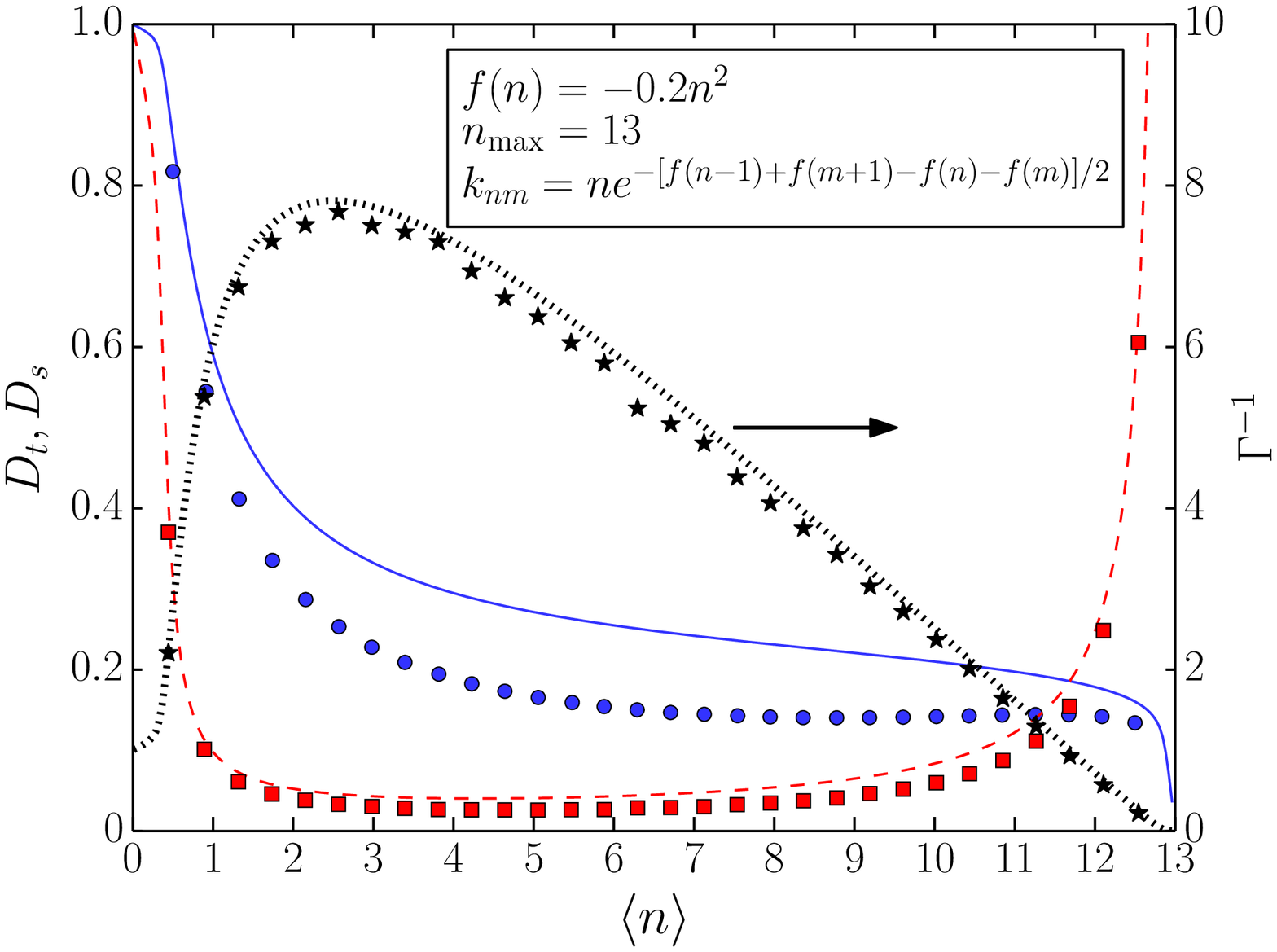}
\label{fig::diff_fnconcave_r2}
 }
\subfloat[][]
{
\includegraphics[width=1\columnwidth]{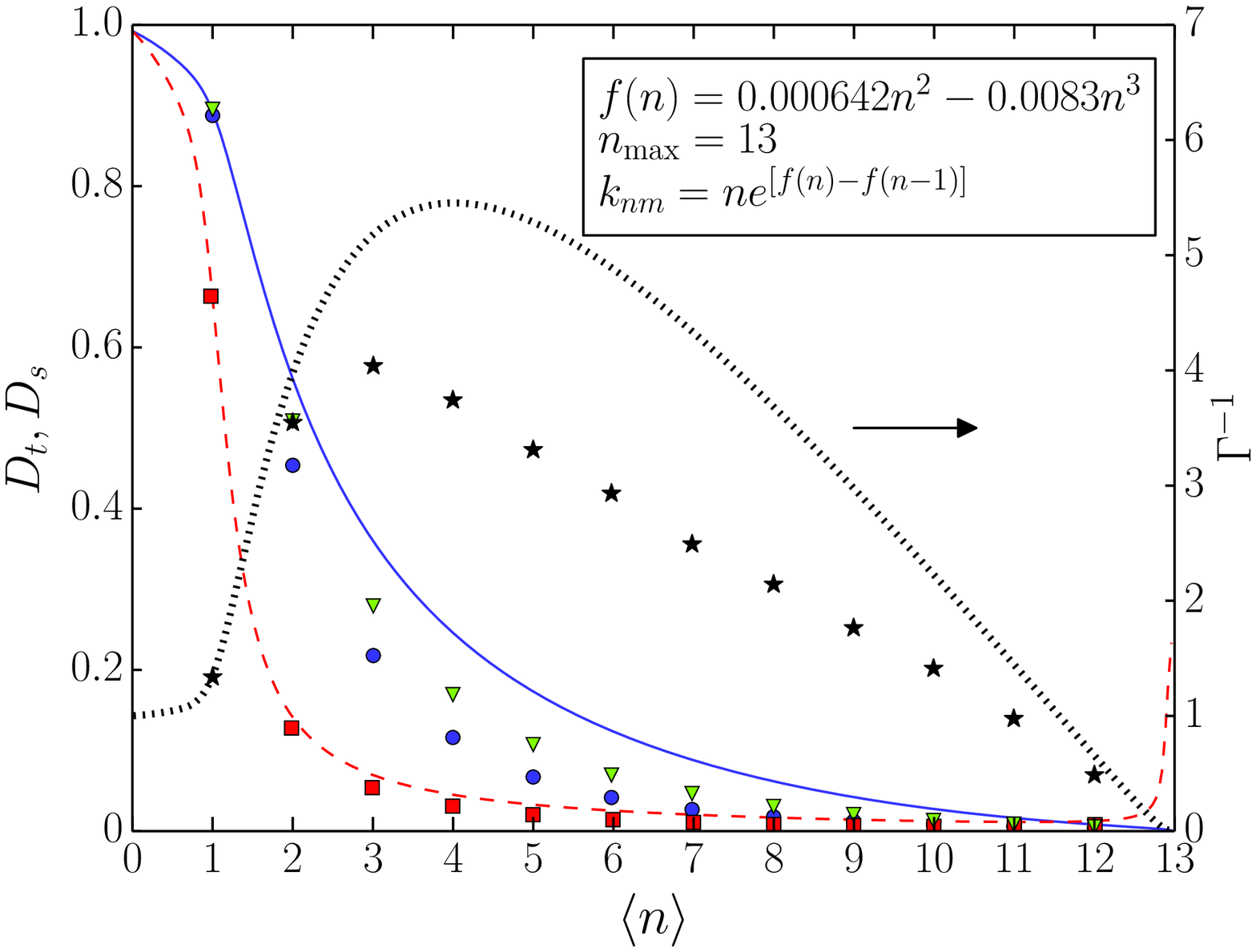}
\label{fig::diff_methanol_r1}
}
\caption{(Color online) $D_s$, $D_t$, and $\Gamma^{-1}$. 
$(a)$ $f(n) = 0$ and $\nmax = 2$; 
$(b)$ $f(n) = 0$ and $\nmax = 13$; 
$(c)$ $f(n) = 0.2 n^2$, $\nmax = 13$ and the rates of Eq.~\eqref{eq::rates2} ;
$(d)$ $f(n) = 0.2 n^2$, $\nmax = 13$ and the rates of Eq.~\eqref{eq::rates1_2}; 
$(e)$ $f(n) = - 0.2 n^2$, $\nmax = 13$ and the rates of Eq.~\eqref{eq::rates2}; 
$(f)$ $f(n) = 0.000 642 n^2 - 0.0083 n^3$, $\nmax = 13$ and the rates of Eq.~\eqref{eq::rates1_2}.}%
\label{fig:diff_fn0_nmax2or13}%
\end{figure*}

\begin{figure}
\centering
\includegraphics[width=1\columnwidth]{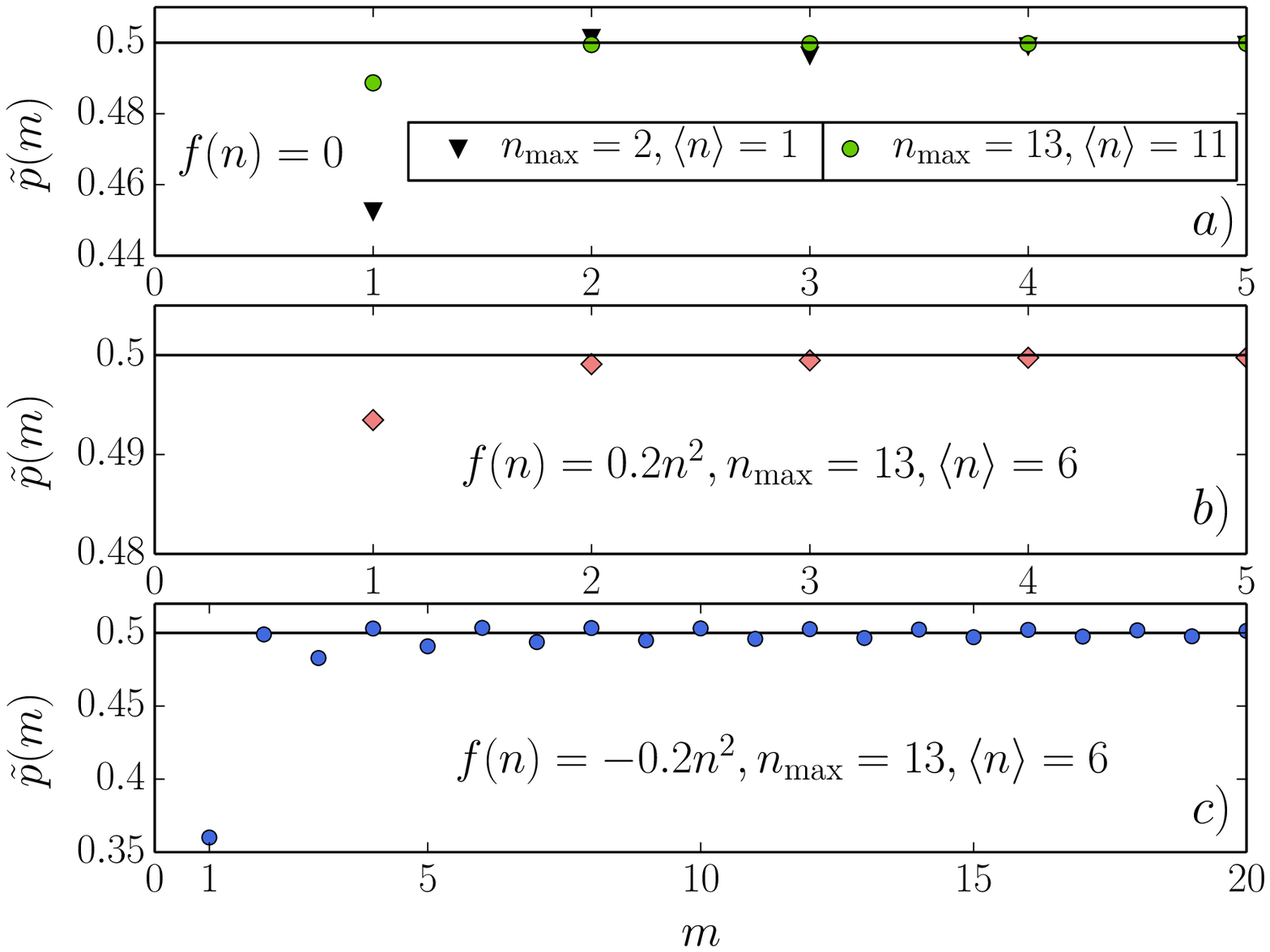}
\caption{(Color online) Single-particle memory $\tilde{p}(m)$ for different interactions and loadings and the rates of Eq.~\eqref{eq::rates2}.}
\label{fig::memory}
\end{figure}

We first discuss the diffusion for $f(n)=0$. In this case the two rates are the same.
Figure \ref{fig::diff_fn0_nmax2} shows the diffusion for $\nmax = 2$. Single-particle correlations are caused by the back-correlation mechanism and lower the self-diffusion significantly. This can be seen in Figure \ref{fig::memory}a, where we plot $\tilde{p}(m)$ at loading $\nav = 1$.
Fig.~\ref{fig::diff_fn0_nmax2} shows that interparticle correlations are positive ($\Gamma^{-1} > D_s/ D_t$), signifying that a particle drags along other particles. This can be understood as follows. Suppose a tagged particle has diffused in a certain direction. The vacancies it leaves behind can be occupied by other particles. In front the other particles have had to ``make way'' for the tagged particle. Both effects cause particles in the environment to diffuse in the same direction as the tagged particle. The transport diffusion is almost equal to the DMF value (different from $\nmax = 1$, for which it is exactly equal). The Maxwell-Stefan diffusion is higher than the self-diffusion and almost equal to the DMF result.

Figure \ref{fig::diff_fn0} shows the diffusion for $f(n) = 0$ and $\nmax = 13$. The diffusive behavior was discussed previously \cite{PRLbecker}. Correlations have a small influence on the self-diffusion, even at high loading. This is because jumps of other particles erase the memory of the environment.
The only type of correlations in the system are back-correlations, which occur at loadings where $p^{\mathrm{eq}}_{\nmax} \neq 0$. For $\nmax = 2$, if the tagged particle jumps from a full cavity, the cavity it jumps to contains at most one other particle. For $\nmax = 13$ at loadings $\nav \approx \nmax$, a particle that jumps from a full cavity will arrive in a cavity containing around 12 other particles. If one of these other particles jumps back, the memory effect of the environment on the tagged particle is lost. 
The back-correlation effect is therefore smaller compared to the case $\nmax = 2$; see $\tilde{p}(m)$ at loading $\nav = 11$ in Fig.~\ref{fig::memory}a.
Similarly to single-particle correlations, interparticle correlations are also small ($\Gamma^{-1} \approx D_s/D_t$). Since the back-correlation mechanism is small, this is what one would expect. In all graphs where $\Gamma^{-1} \approx D_s/D_t$ one has that $D_{\mathrm{ms}} \approx D_s$; we do not plot the MS diffusion for these cases.

We now discuss the diffusion for $f(n) = 0.2 n^2$ and $\nmax = 13$.
Figure \ref{fig::diff_fnconvex_r1} shows the diffusion for the rates of Eq.~\eqref{eq::rates1_2}. The self-diffusion shows an increasing trend with increasing concentration and a decrease near $\nav \approx \nmax$. This is typical behavior observed in MD simulations \cite{PRLbeerdsen2006,JPCBbeerdsen2006}. Similar behavior is obtained for repulsive particles in other lattice models \cite{PCCPkrishna2013,JCPBdemontis2008,JCPpazzona20092}. For convex $f(n)$'s with the rates of Eq.~\eqref{eq::rates1_2} this behavior always occurs, as can be understood as follows.
The difference $\left[ f(n-1) + f_{\mathrm{TS}} \right] - f(n)$ measures the change in interaction free energy when a particle moves to the TS ($ f_{\mathrm{TS}}$ is the interaction free energy of a particle in the TS, cf.~Appendix \ref{app:tstcalc}). Since $f(n)$ is convex the difference $f(n) - f(n-1)$ grows with increasing $n$: It becomes easier to jump to the TS for higher loadings, increasing the diffusion. This interpretation of such behavior is well known \cite{PRLbeerdsen2004,CSRkrishna2012,PCCPkrishna2013}.
As long as the system does not feel that there is an $\nmax$ ($p^{\mathrm{eq}}_{n_{\mathrm{max}}} \approx 0$) the dynamics is a ZRP and the DMF solution is exact; see Sec.~\ref{sec::DMFA}. When the presence of $\nmax$ is felt there are correlations because of the back-correlation mechanism. We can conclude that the rates of Eq.~\eqref{eq::rates1_2} provide the correct qualitative behavior for repulsive particles, as observed in MD simulations \cite{PRLbeerdsen2006,JPCBbeerdsen2006,PCCPkrishna2013}.

\begin{figure}%
\centering
\subfloat[][]{\includegraphics[width=0.9\columnwidth]{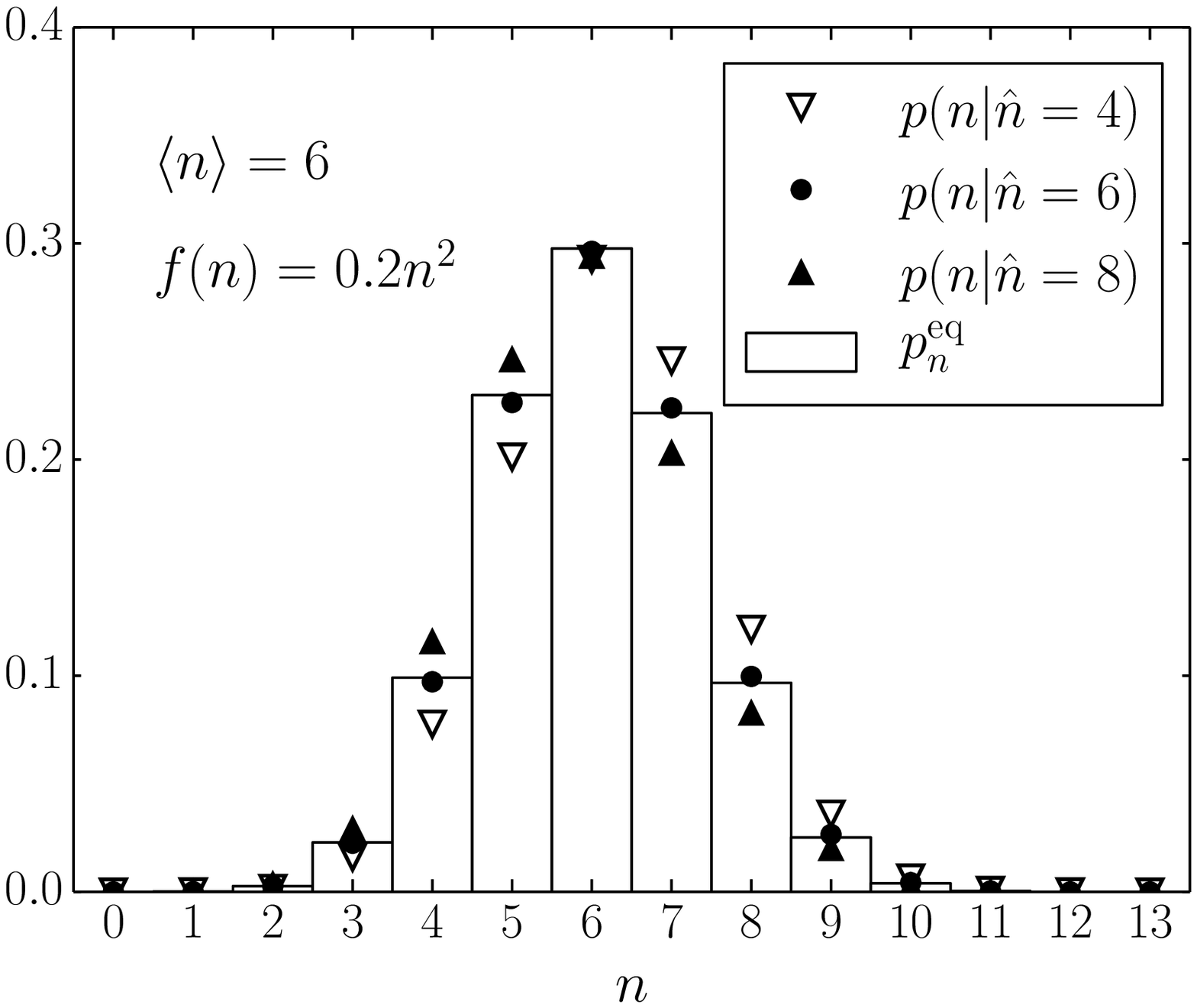}
\label{fig::corr_fnconvex_r2}
 }
\\
\subfloat[][]
{
\includegraphics[width=0.9\columnwidth]{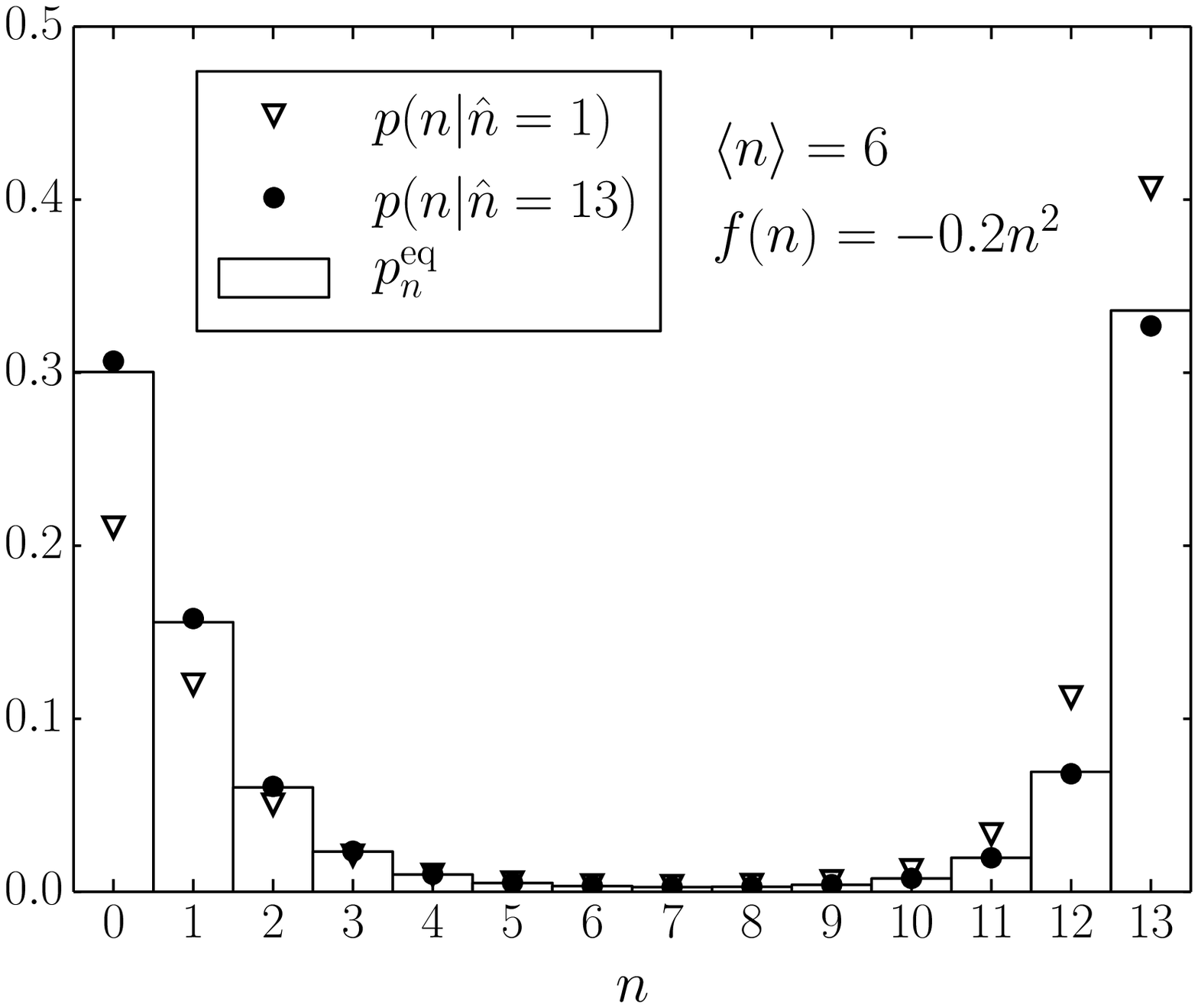}
\label{fig::corr_fnconcave_r2}
}
\caption{$p^{\mathrm{eq}}_n$ and $p(n | \hat{n})$ for different $\hat{n}$ at loading $\nav = 6$, $\nmax = 13$, and the rates of Eq.~\eqref{eq::rates2} for $(a)$ $f(n) = 0.2 n^2$ and $(b)$ $f(n) = - 0.2 n^2$.}%
\label{fig:memorypn}%
\end{figure}

Figure \ref{fig::diff_fnconvex_r2} shows the diffusion for $f(n) = 0.2 n^2$, $\nmax = 13$, and the rates of Eq.~\eqref{eq::rates2}. A close-up of the self-diffusion is shown in Fig.~\ref{fig::diff_fnconvex_dimension}. The diffusive behavior was discussed previously \cite{PRLbecker} and qualitatively differs from Fig~\ref{fig::diff_fnconvex_r1}. It was noted that correlations have a small influence on the diffusion. 
Because the rates depend on the number of particles in both cavities there are correlations caused by the interaction, besides the back-correlation mechanism. 
This can be seen by comparing the self-diffusion (Fig.~\ref{fig::diff_fnconvex_dimension}) with the self-diffusion for $f(n)=0$ and $\nmax = 13$ (Fig.~\ref{fig::diff_fn0}). While correlations only have an influence for $f(n)=0$ if the presence of $\nmax$ is felt, correlations are also present at loadings where $\nmax$ is not felt for $f(n) = 0.2 n^2$.
We investigate correlations at loading $\nav = 6$. The probability to be full is then negligible [see $p^{\mathrm{eq}}_n$ in Fig.~\ref{fig::corr_fnconvex_r2}], and all correlation effects are caused by the interaction.
Single-particle correlations due to the interaction can be understood as follows. 
Consider two cavities containing six particles. If the tagged particle hops, $(6,6) \rightarrow (5,7)$, the average number of particles in the cavity it came from is smaller compared the average of $p^{\mathrm{eq}}_n$. Because the free energy favors a homogeneous density distribution, this increases the rate at which the particle jumps back, lowering the self-diffusion compared to the DMF approximation. If the tagged particle makes the jump $(8,4) \rightarrow (7,5)$, the rate to jump back is smaller compared to the DMF approximation, thereby enhancing the diffusion. 
$\tilde{p}(m) \approx 1/2$ for all $m > 1$ and is slightly smaller for $m = 1$; see Fig.~\ref{fig::memory}b. Memory effects are small \textit{on average}: The self-diffusion is around 2.6 \% lower than the DMF value. $p(n | \hat{n})$ for different $\hat{n}$ and $p^{\mathrm{eq}}_n$ are shown in Fig.~\ref{fig::corr_fnconvex_r2}. There is a clear difference between $p(n | \hat{n})$ and $p^{\mathrm{eq}}_n$ for $\hat{n} \neq \nav$. There are two reasons why the effect of correlations on the diffusion is small. Because there are on average 6 particles per cavity, jumps of other particles tend to erase the memory effect of the environment, as discussed previously. The effect on the diffusion is further diminished because correlations contribute both positively and negatively. They therefore partly cancel each other. 

Figure \ref{fig::diff_fnconcave_r2} shows the diffusion for $f(n) = -0.2 n^2$, $\nmax = 13$, and the rates of Eq.~\eqref{eq::rates2}. We refer to Ref.~\cite{PRLbecker} for a discussion of the diffusive behavior. There are strong memory effects; see $\tilde{p}(m)$ for $\nav = 6$ in Fig.~\ref{fig::memory}c. Not only is $\tilde{p}(1)$ much smaller than 1/2, the memory effect is also long lived. These strong correlations are caused by the clustering of the particles. An example of a strongly correlated event is when the tagged particle jumps from a full to an empty cavity. The probability to jump back is then large; see the difference between $p^{\mathrm{eq}}_n$ and $p(12 |\hat{n} = 1)$ at $\nav = 6$ in Fig.~\ref{fig::corr_fnconcave_r2}. For this event there are no other particles in the cavity where the tagged particle jumps to whose presence could decrease the memory effect. Even though correlations have a strong effect on the self- and transport diffusion, interparticle correlations are small ($\Gamma^{-1} \approx D_s / D_t$). To understand why particles do not drag along other particles, we examine the dynamics more carefully. For all $n \leq \nmax$ one has that $k_{n,n-1} = n k_{10}$, which is the same rate as for $f(n) = 0$. In other words, particle exchange between cavities that are both almost full or both almost empty follows a dynamics similar to the situation for noninteracting particles. In this case, particles do not drag along other particles; see Fig.~\ref{fig::diff_fn0}. As can be seen from $p^{\mathrm{eq}}_n$ (Figs.~\ref{fig::histogram_fnconcave} and \ref{fig::corr_fnconcave_r2}), this type of transition occurs a lot. The other type of transition that often occurs is a particle jump from a full to an empty cavity. These are strongly correlated events, but they influence only the diffusion of a \textit{single} particle.

We now discuss the diffusion for the concave free energy $f(n) = 0.000 642 n^2 - 0.0083 n^3$ with $\nmax = 13$, which was obtained by fitting the analytical $\Gamma$ from our model with the experimental $\Gamma$ of methanol in ZIF-8 \cite{PRLchmelik2010}. The experimental $\Gamma$ is calculated from the experimentally measured adsorption isotherm. However, $f(n)$ does not specify the type of rates that should be used, as discussed in Sec.~\ref{sec::transrates}. For the rates of Eq.~\eqref{eq::rates2} a good agreement with the experimental diffusion data was found; see Fig.~3 in Ref.~\cite{PRLbecker}. In contrast, Fig.~\ref{fig::diff_methanol_r1} shows the diffusion for the same parameters, with the rates of Eq.~\eqref{eq::rates1_2}. The diffusive behavior qualitatively differs: both self- and transport diffusion become much smaller for high loadings, and $\Gamma^{-1} > D_s / D_t$. This is in contrast to the experiments where $\Gamma^{-1} \approx D_s / D_t$ \cite{PRLchmelik2010}, which is also reproduced by the rates of Eq.~\eqref{eq::rates2} \cite{PRLbecker}. We therefore conclude that for clustering particles the rates of Eq.~\eqref{eq::rates2} give the correct qualitative behavior of the diffusion. This is further supported by the discussion of the calculation of the transition rates in Sec.~\ref{sec::transrates} and Appendix \ref{app:tstcalc}. Correlations have a strong effect on the diffusion, cf.~Fig.~\ref{fig::diff_methanol_r1}. Because the rates only depend on the number of particles in the cavity of the tagged particle, these correlations are caused by the back-correlation mechanism. Since the probability to be full is non-negligible even at low loadings, this is not surprising. $\Gamma^{-1} > D_s / D_t$, i.e., interparticle correlations are positive. Since the back-correlation mechanism has a strong impact on the self-diffusion one expects significant inter-particle correlations, as discussed for $f(n) = 0$ and $\nmax = 2$. 
The Maxwell-Stefan diffusion is higher than the self-diffusion but still significantly smaller than the DMF result.
Note that because $p^{\mathrm{eq}}_{\nmax} \neq 0$ even at low loadings, the dynamics can never be approximated by a ZRP.

\begin{figure}%
\centering
\subfloat[][]{\includegraphics[width=1\columnwidth]{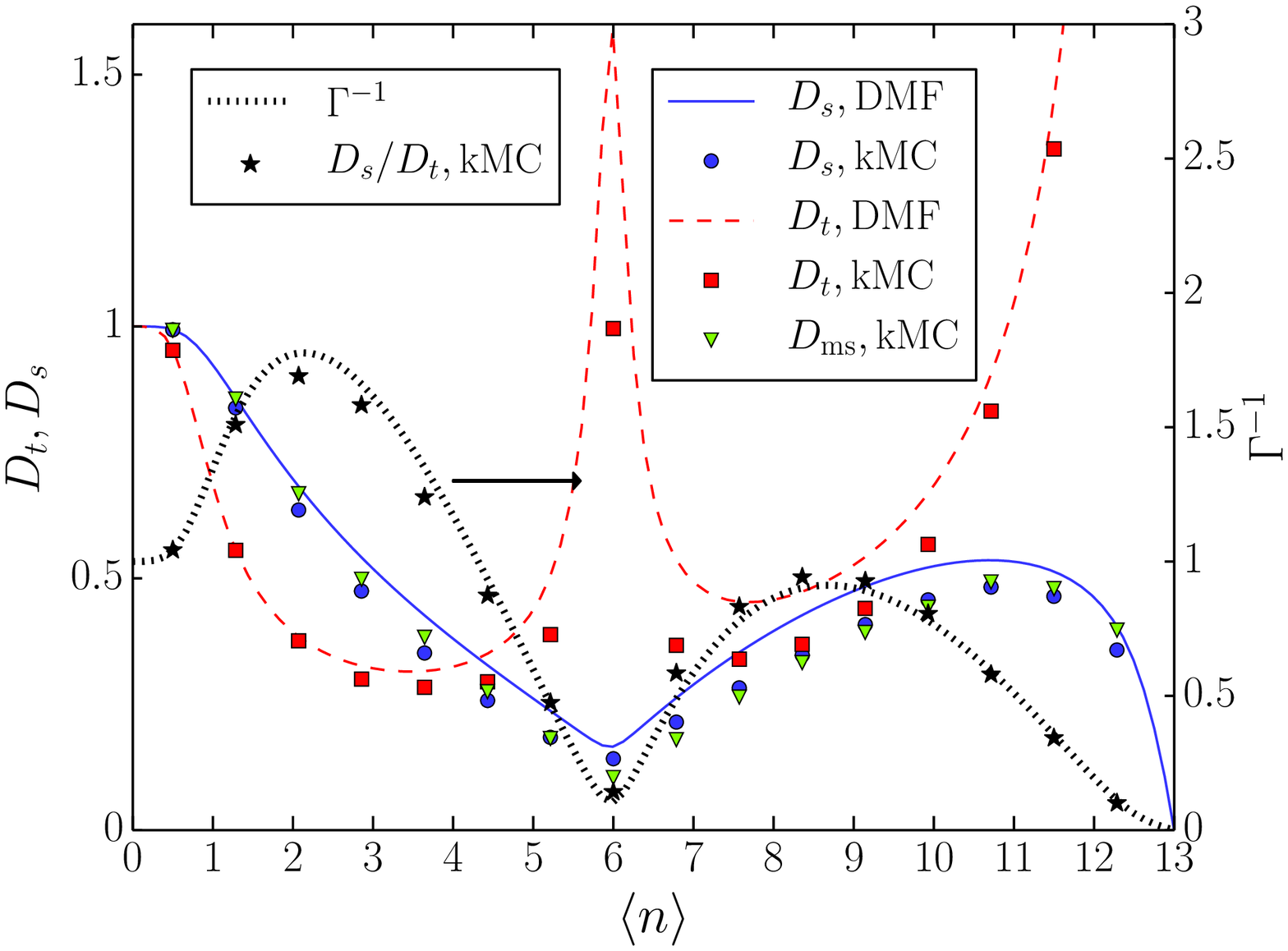}
\label{fig::diff_fnspecial}
}
\\ 
\subfloat[][]
{
\includegraphics[width=1\columnwidth]{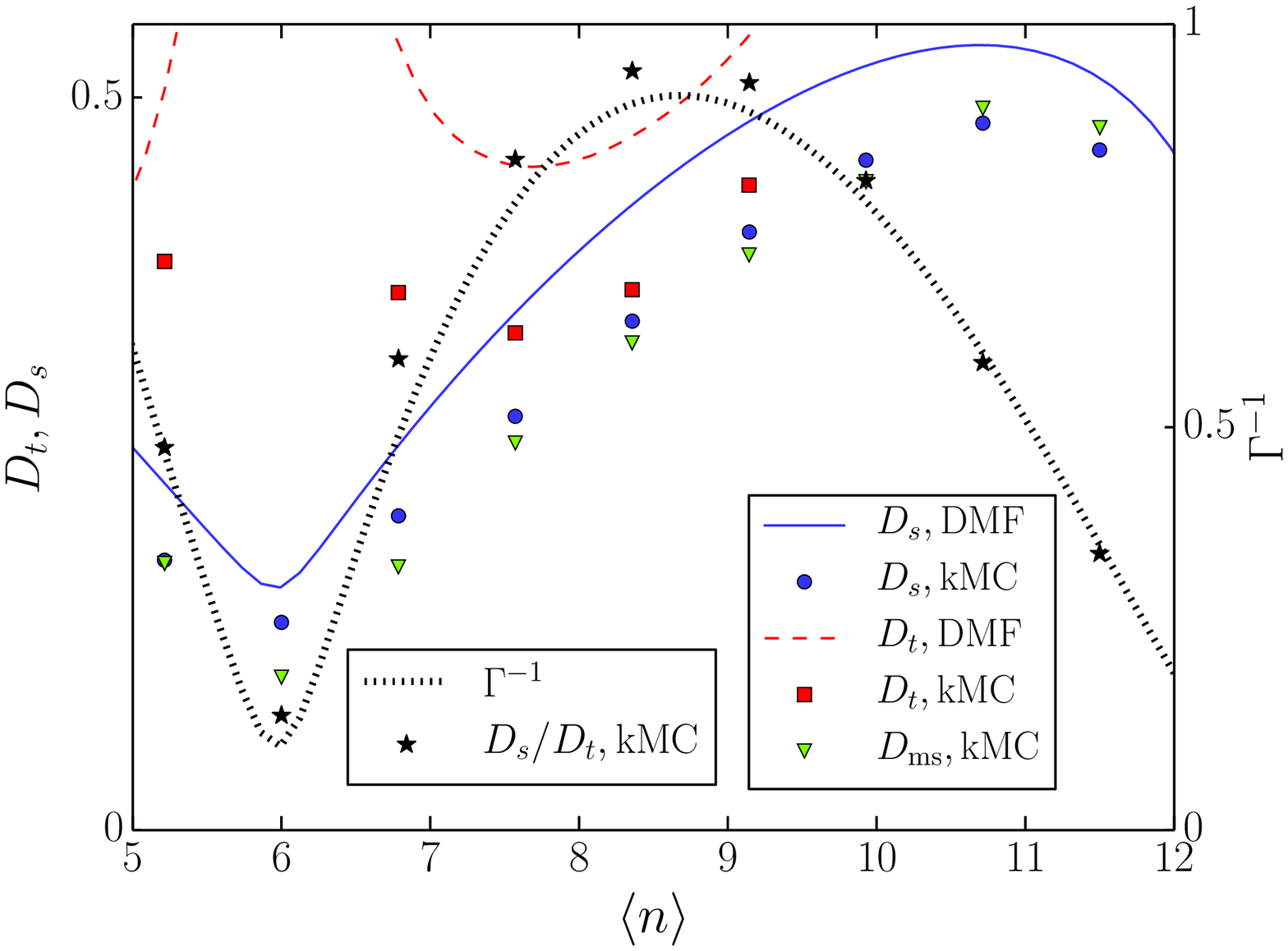}
\label{fig::diff_fnspecial_closeup}
}
\\ 
\subfloat[][]
{
\includegraphics[width=1\columnwidth]{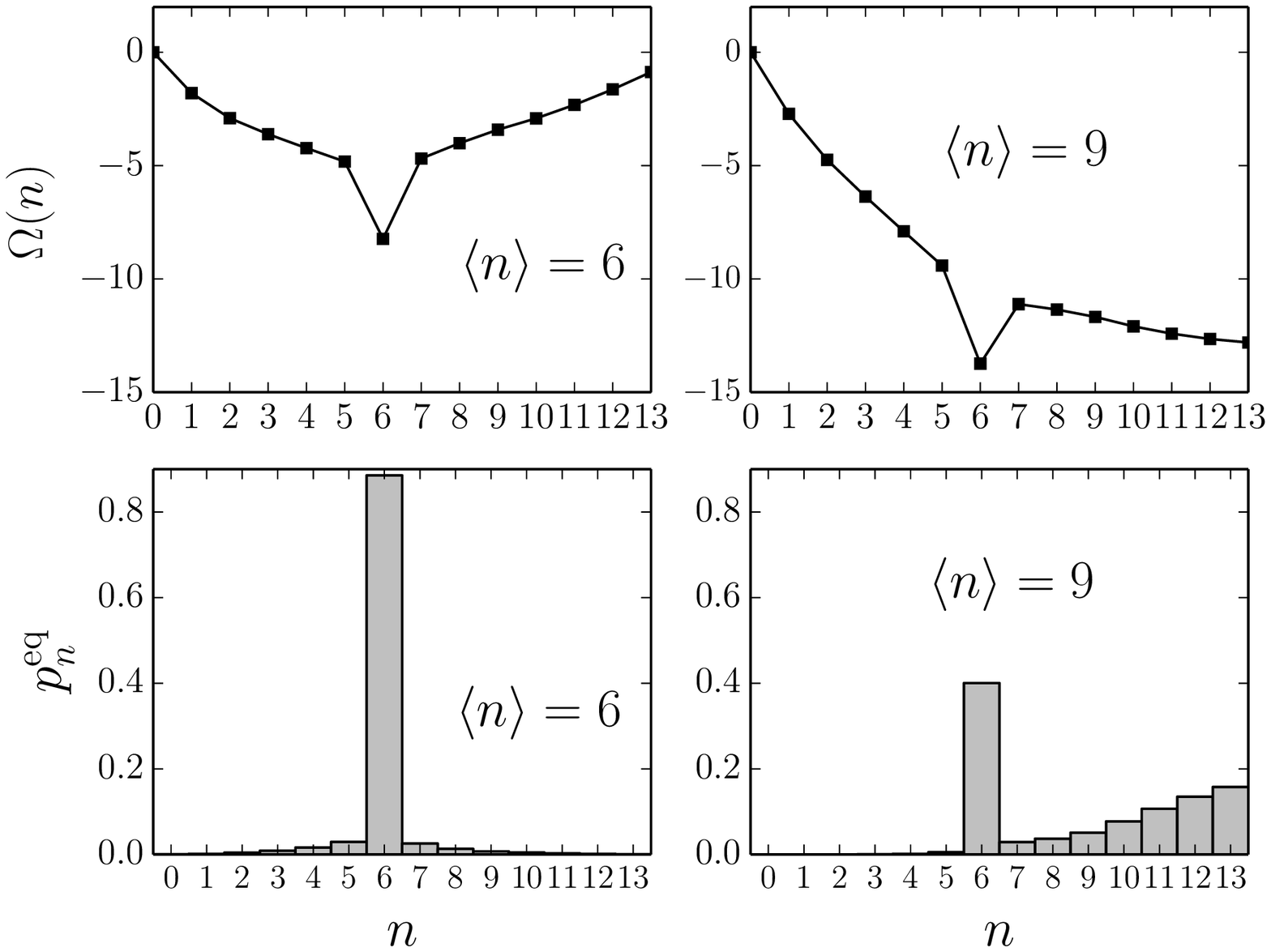}
\label{fig::Gn_fnspecial}
}
\caption{(Color online) $f(n)$ of Table \ref{tab::1}, $\nmax = 13$, and the rates of Eq.~\eqref{eq::rates2}. $(a)$ $D_s$, $D_t$, and $\Gamma^{-1}$. $(b)$ Close-up of diffusion curves for $5 \leq \nav \leq 12$.  $(c)$ $\Omega(n)$ and $p^{\mathrm{eq}}_n$ at $\nav = 6$ and $\nav = 9$.}%
\label{fig:fnspecial_all}%
\end{figure}

\begin{table}[h]
\begin{center}
\begin{tabular}{  l | c|c|c|c|c|c|c|c|c|c|c|c|c|c }
$n$ & 0 & 1 & 2 & 3 & 4 & 5 & 6 & 7 & 8 & 9 & 10 & 11 & 12 & 13 \\ \hline
$f(n)$ & 0 & 0 & 0 & 0 & -0.2 & -0.6 & -4.0 & -0.6 & -0.2 & 0 & 0 & 0 & 0 & 0 \\
\end{tabular}
\caption{\label{tab::1} $f(n)$ that switches among concave, convex, and concave.}
\end{center}
\vspace{-0.6cm}
\end{table}

Figure \ref{fig::diff_fnspecial} shows the diffusion for $\nmax = 13$, the rates of Eq.~\eqref{eq::rates2}, and the $f(n)$ from Table \ref{tab::1}. We study this interaction because it switches among concave, convex, and concave. This in contrast to the previous interactions, which are concave, convex, or constant over the whole concentration range. 
We again refer to Ref.~\cite{PRLbecker} for a discussion of the diffusion. We focus here on the fact that $\Gamma^{-1} < D_s/D_t$ for $6 \leq \nav \leq 10$ [see Fig.~\ref{fig::diff_fnspecial_closeup}], implying negative interparticle correlations. The grand potential $\Omega(n)$ and $p^{\mathrm{eq}}_n$ at loadings $\nav = 6$ and $\nav = 9$ are plotted in Fig.~\ref{fig::Gn_fnspecial}. The crucial property to obtain $\Gamma^{-1} < D_s/D_t$ is that the cavity occupation $n = 6$ is very stable, while all other occupations around it are not. 
Consider a tagged particle that has diffused in a certain direction, in a system at loading $\nav = 6$. In this case almost all cavities contain six particles. When the tagged particle jumps to a neighboring cavity, $(6,6) \rightarrow (5,7)$, it immediately pushes one of the other particles in its new cavity to the cavity it came from to restore the situation where every cavity has six particles. A particle that has diffused in a certain direction therefore pushes other particles in the opposite direction. The interparticle correlation term in Eq.~\eqref{eq::dsdtintpartcorr} is then negative and $\Gamma^{-1} < D_s/D_t$. 
In the theory of Maxwell-Stefan diffusion this means that $1/ D_{\mathrm{cor}} < 0$ and $D_{\mathrm{ms}} < D_s$, as shown in Fig.~\ref{fig::diff_fnspecial_closeup}. We are unaware of any previous studies where $D_{\mathrm{ms}} < D_s$ was found (or at least explicitly mentioned). Indeed, it is often assumed that the self-diffusion is always higher than the Maxwell-Stefan diffusion \cite{JPCCkrishna2009}.

\begin{figure}
\centering
\includegraphics[width=0.9\columnwidth]{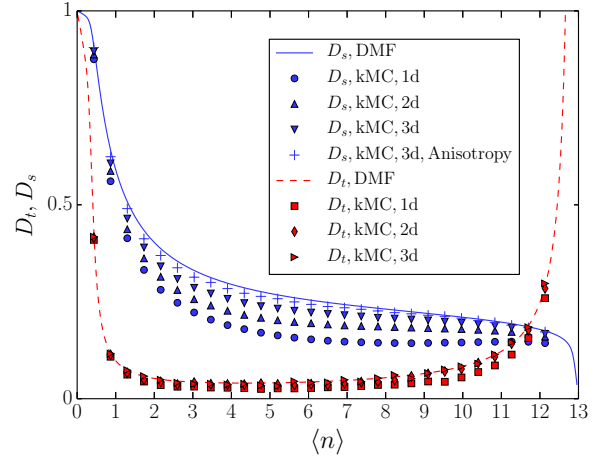}
\caption{(Color online) $D_s$ and $D_t$ for $f(n) = - 0.2 n^2$, $\nmax = 13$, and the rates of Eq.~\eqref{eq::rates2} for dimensions 1, 2, and 3.}
\label{fig::diff_fnconcave_dimension}
\end{figure}
\begin{figure}
\centering
\includegraphics[width=0.9\columnwidth]{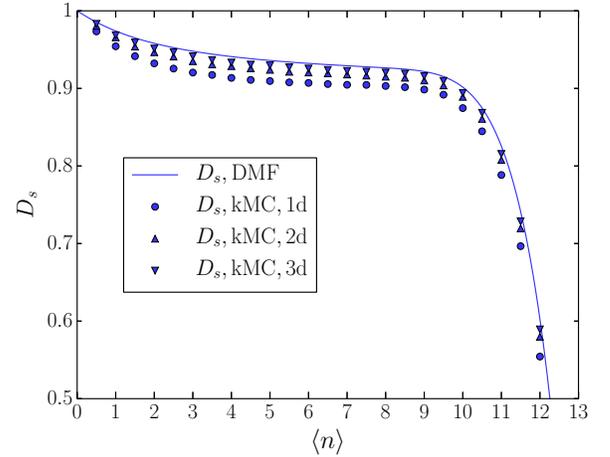}
\caption{(Color online) Self-diffusion for $f(n) = 0.2 n^2$, $\nmax = 13$, and the rates of Eq.~\eqref{eq::rates2} for dimensions 1, 2, and 3.}
\label{fig::diff_fnconvex_dimension}
\end{figure}

We now discuss the diffusion in two- and three-dimensional systems. The two-dimensional lattice has a square geometry and the three-dimensional lattice is cubic. For these geometries the DMF results Eqs.~\eqref{eq::dsnocorr} and \eqref{eq::dtnocorr} are equal to the one-dimensional case; see Sec.~\ref{sec::DMFA}. 
In Fig.~\ref{fig::diff_fnconcave_dimension} we plot the self- and transport diffusion for $f(n) = - 0.2 n^2$, $\nmax = 13$, and the rates of Eq.~\eqref{eq::rates2} in one, two, and three dimensions. The qualitative behavior stays the same. The effect of correlations decreases with increasing dimension, which can be  understood as follows. If a tagged particle has jumped, most of the memory of the environment comes from the cavity from which it came. The relative influence of this cavity decreases if there are more cavities connected to the cavity of the tagged particle. One therefore expects the effect of correlations to decrease proportionally to the number of neighbors of each cavity. Also plotted is the diffusion in the $x$ direction in a three-dimensional system, where the rates in the $y$ and $z$ directions are $10$ times faster than the rates in the $x$ direction (i.e., there is anisotropy in the dynamics). The effect of correlations is further reduced: If a particle jumps in the $x$ direction, the memory of the environment is erased faster by particles jumping the $y$ and $z$ directions. In Fig.~\ref{fig::diff_fnconvex_dimension} we plot the self-diffusion for $f(n) = 0.2 n^2$, $\nmax = 13$, and the rates of Eq.~\eqref{eq::rates2} in one, two, and three dimensions. Also here the qualitative behavior stays the same, with the effect of  correlations decreasing with higher dimensionality. The effect of correlations on the transport diffusion is already negligible in one dimension, so we do not plot the data for higher dimensions.

\section{\label{sec::assumptions}Correspondence with simulations and experiments}
In Ref.~\cite{PRLbecker} quantitative agreement with experimental results of methanol and ethanol diffusion in MOF ZIF-8 was found. In this case the particles undergo strong (attractive) interactions via hydrogen bonding \cite{LANGMUIRkrishnaHbond}.
The behavior of both the self- and transport diffusion was reproduced correctly over the whole concentration range. This was achieved for an interaction free energy that correctly reproduced the experimentally measured thermodynamic factor. We now examine the assumptions underlying our model and discuss when agreement with experiments and molecular dynamics simulations can be expected.

We assume that particles in neighboring cavities do not interact with each other. The only exception is when a particle is in the transition state, in which case it can interact with particles in both cavities, as discussed in Sec.~\ref{sec::transrates}. The same assumption has been made when modeling the behavior of adsorption isotherms; see Sect.~\ref{sec::adsiso}. A good agreement with experiments can be achieved for apolar molecules at low and medium loadings \cite{BOOKruthven}. For high loadings and polar molecules this assumption is generally not quantitatively correct. 
The importance of intercavity interactions on equilibrium properties was investigated using simulations in Refs.~\cite{CESTunca2003,JPCCpazzona2012}. It was found that at high loadings this interaction is in general non-negligible. At low temperature intercavity interactions can be important at low loadings, cf.~Ref.~\cite{JPCCpazzona2012}. In Ref.~\cite{PRLbeerdsen2004} the average jump rate of a particle between two cavities ($\hat{k}_{\mathrm{av}}$) was calculated numerically using dynamically corrected TST. It was found that if all other cavities (differing from cavities A and B as defined in Sec.~\ref{sec::transrates}) connected to the cavity of the particle that jumps are closed off, the calculated self-diffusion can differ by as much as 60 \%. 
From these results it is clear that agreement with experiments can, in general, only be expected for low and medium loadings. The agreement with experiment was found for strongly interacting particles, using the rates of Eq.~\eqref{eq::rates2}. For these rates, the particle in the transition state interacts with the particles in both cavities. Hence, the most important intercavity interaction, when one of the particles is in the transition state, is taken into account in the dynamics.

We did not include a dynamical correction factor in the rates, i.e., we assumed that all particles crossing the TS equilibrate in cavity B (see the discussion in Appendix \ref{app:tstcalc}). A quantitative influence of the correction factor was found for particles feeling repulsive interactions \cite{PRLbeerdsen2004,JPCBbeerdsen2006}. This is understandable for repulsive interactions, because particles in cavity B tend to push back the particle that jumps. Agreement was found with experiments of clustering particles, which are attractive. It can be expected that recrossing of the TS on short time scales are of less importance in this case, because a particle that has crossed the TS is attracted by the other particles in cavity B.

Flexibility of the material can have an influence, as was found for ethane diffusion in MOF ZIF-8 \cite{MMMchokbunpiam2013}. It remains an open question if it is important for ethanol and methanol diffusion.
We note that cavity windows whose size depends on the loading, as found in Ref.~\cite{MMMchokbunpiam2013}, can be accounted for by making the interaction free energy of the TS depend on the number of particles: $f_{\mathrm{TS}} = f_{\mathrm{TS}}(n)$.

From the above discussion one can conclude that in general a qualitative agreement can be expected with experimental systems.
For the case of clustering particles, our model seems to allow for a quantitative agreement of both dynamical and equilibrium properties over the whole concentration range. Both the free energy $F(n)$ and the rates $k_{nm}$ can be determined numerically using MD simulations \cite{CESTunca2003,PCCPabou,JACSsholl}. The quality of our assumptions and choice of rates could be verified using these techniques. Such a study would also be of interest to investigate memory effects. Beerdsen, Dubbeldam, and Smit have studied diffusion in microporous materials using dynamically corrected TST \cite{PRLbeerdsen2004,JCPdubbeldam2005,PRLbeerdsen2006}. In their work it is assumed that, after the particle has equilibrated in the cavity it has jumped to, memory effects are negligible. Abouelnasr and Smit presented a study where these memory effect are included \cite{PCCPabou} for a system showing behavior similar to that in Fig.~\ref{fig::diff_fnconvex_r1}. In this case memory effects in the environment can be expected to be negligible, as was found in Ref.~\cite{PCCPabou}. For clustering particles this memory effect is, however, much stronger, especially in a one-dimensional system, cf.~Fig.~\ref{fig::diff_fnconcave_r2}.
\section{\label{sec::conclude}Conclusion}
To conclude, we have studied a lattice model of interacting particles. It is assumed that particles are in equilibrium on the lattice sites (also called cavities). The equilibrium free energy $F(n)$ of $n$ particles in a cavity then describes all the interactions.
The equilibrium properties, such as the adsorption isotherm and the thermodynamic factor, only depend on $F(n)$ and the maximum number of particles in a cavity $\nmax$. The qualitative behavior observed in experiments and molecular dynamics simulations is reproduced in the model, while allowing for a simple physical interpretation. Different forms of the transition rates are calculated using transition-state theory. The qualitative diffusive behavior of both clustering (attractive) and repulsive particles is reproduced in our model, where both cases require the use of different rates.
Memory effects in the environment lead to correlations in the dynamics. If these correlations are neglected analytical expressions for the self- and transport diffusion can be derived for all interactions and loadings. Correlations are studied by comparing these expressions with the diffusion coefficients obtained from kinetic Monte Carlo simulations. 
For certain interactions the self-diffusion can exceed the Maxwell-Stefan diffusion. We are unaware of any previous studies that have found this. The higher the dimension of the system the smaller the effect of correlations on the diffusion. This is because the number of neighboring cavities grows with increasing dimension, thereby lowering the memory effect of the environment. 
The assumption that particles in different cavities do not interact is in general only valid at low and medium concentrations for particles that are not strongly interacting.
The choice of rates has a strong influence on the diffusion. For the reproduction of experimental data of clustering particles, it is important that the dynamics depends on the free energy of both cavities participating in the particle jump. 

\begin{acknowledgments}
This work was supported by the Flemish Science Foundation (Fonds Wetenschappelijk Onderzoek), Project No.~G038811N. The computational resources and services used in this work were provided by the VSC (Flemish Supercomputer Center), funded by the Hercules Foundation and the Flemish Government -- department EWI.
\end{acknowledgments}
\appendix
\section{Transition-state-theory calculations}\label{app:tstcalc}

Because the window separating the two cavities is a perfect choice for the transition state, the system under study is ideally suited for a TST calculation; see, e.g., Refs.~\cite{TSTruthven,TSTkarger,JPCjune,IRPCauerbach,JCPTunca1999,JPCBTunca2002}. 
We use the expression given by Tunca and Ford \cite{JPCBTunca2002}: 
\begin{equation}\label{eq::rateTST}
k^{\mathrm{TST}}_{n m} = \left( 2 \pi M \beta \right)^{-1/2}  \frac{S}{V} n \frac{z(n-1,1,m)}{z(n) z(m)},
\end{equation}
where $M$ is the mass of one particle, $S$ the area of the TS surface, and $V$ the volume of the cavity. $z(n,1,m)$ is the configurational integral with a particle in the TS:
\begin{align}
z(n,1,m) =
\frac{1}{S V^{n+m}} \int_{S} \int_{V_A} \int_{V_B} d \mathbf{r}^{\mathrm{TS}}  d \mathbf{r}^{\mathrm{A}} d\mathbf{r}^{\mathrm{B}}  e^{-\beta U_{\mathrm{tot}}} \label{eq::confintTS}.
\end{align}
The labels $\mathbf{r}^{\mathrm{TS}}$, $\mathbf{r}^{\mathrm{A}}$, and $\mathbf{r}^{\mathrm{B}}$ denote the positions of all the particles in, respectively, the TS, cavity $A$, and cavity $B$. 
$U_{\mathrm{tot}}$ is the total interaction energy of the particles in the TS and both cavities. It is assumed that the TS can hold at most one particle. The TST rate of Eq.~\eqref{eq::rateTST} has a simple physical interpretation: The second term gives the probability that a particle from cavity $A$ is in the TS, while the first term $\left( 2 \pi M \beta \right)^{-1/2}$ is the average velocity towards cavity $B$ of a particle in the TS. Their product gives the rate at which a particle jumps from cavity $A$ to cavity $B$. Note that these rates always satisfy local detailed balance. 

The total interaction energy can be written as follows:
\begin{equation}
U_{\mathrm{tot}}=U_{\mathrm{TS}}(\mathbf{r}^{\mathrm{TS}}) 
+ U(\mathbf{r}^{\mathrm{A}}) + V(\mathbf{r}^{\mathrm{A}}, \mathbf{r}^{\mathrm{TS}}) + U(\mathbf{r}^{\mathrm{B}}) + V(\mathbf{r}^{\mathrm{B}}, \mathbf{r}^{\mathrm{TS}}). \label{eq::Utot}
\end{equation}
$U_{\mathrm{TS}}(\mathbf{r}^{\mathrm{TS}})$ is the energy of the particle in the TS due to interactions with the cavity wall. The interaction energy in cavity $A$ is equal to $U(\mathbf{r}^{\mathrm{A}}) + V(\mathbf{r}^{\mathrm{A}}, \mathbf{r}^{\mathrm{TS}})$. $U(\mathbf{r}^{\mathrm{A}})$ is the same function as in Eq.~\eqref{eq::confint}, i.e., the total interaction energy in cavity $A$ if there is no particle in the TS. $V(\mathbf{r}^{\mathrm{A}}, \mathbf{r}^{\mathrm{TS}})$ is the contribution to the interaction energy of cavity $A$ caused by the particle in the TS. 

We now derive the transition rate of Eq.~\eqref{eq::rates1_2}. Since there are no long-range interactions, one can make the approximation that the particle in the TS does not influence the particles in the cavities: $V(\mathbf{r}^{\mathrm{A}}, \mathbf{r}^{\mathrm{TS}}) = V(\mathbf{r}^{\mathrm{B}}, \mathbf{r}^{\mathrm{TS}}) = 0$. Using Eq.~\eqref{eq::Utot} one finds that the configuration integral Eq.~\eqref{eq::confintTS} can be written as $z(n,1,m) = z_{\mathrm{TS}} z(n) z(m)$, with the definition:
\begin{align}
z_{\mathrm{TS}} = \frac{1}{S} \int_{S} d \mathbf{r}^{\mathrm{TS}} e^{- \beta U_{\mathrm{TS}}(\mathbf{r}^{\mathrm{TS}}) }. \label{eq::zTS}
\end{align}
Writing $f_{\mathrm{TS}} = - k T \ln z_{\mathrm{TS}}$ one finds the transition rate:
\begin{equation}
k_{nm} = \left( 2 \pi M \beta \right)^{-1/2} \frac{S}{V} e^{- \beta f_{\mathrm{TS}} } n e^{\beta \left[ f(n) - f(n-1) \right]}. \label{eq::rates1_app}
\end{equation}
Rewriting this as a function of $k_{10}$ gives Eq.~\eqref{eq::rates1_2}.

For long-range particle interactions the assumption that $V(\mathbf{r}, \mathbf{r}^{\mathrm{TS}}) = 0$ no longer holds true. The specific form of $V(\mathbf{r}, \mathbf{r}^{\mathrm{TS}})$ depends on the interparticle interactions. An analytical calculation of $z(n,1,m)$ is difficult in this case. It also can no longer be expected that $z(n,1,m)$ can be written as a function of $f(n)$. We can, however, put upper and lower bounds on $f(n,1,m) = - k T \ln z(n,1,m)$ as a function of $f(n)$. Consider $f(n,m | \mathrm{TS}) = f(n,1,m) - f_{\mathrm{TS}}$, with $f_{\mathrm{TS}}$ the interaction free energy of a particle in the TS that has no interaction with particles in the cavities: $f_{\mathrm{TS}} = - kT \ln z(0,1,0)$. All interparticle interactions are then included in $f(n,m | \mathrm{TS})$. This function must lie in between:
\begin{equation}
f(n) + f(m) \leq f(n,m | \mathrm{TS}) \leq f(n+1) + f(m+1). 
\end{equation}
The lower bound becomes an equality if the particle in the TS does not interact with the particles in the cavities, in which case one finds Eq.~\eqref{eq::rates1_app} for $k_{nm}$. This corresponds to $c=1$ in Eq.~\eqref{eq::GDBrate}. The upper bound becomes an equality if the particle in the TS interacts with the particles in the cavities in exactly the same way as if it was located in the cavities. This gives the rates with $c=0$ in Eq.~\eqref{eq::GDBrate}.
The rates of Eq.~\eqref{eq::rates2} are found for the choice $f(n-1,m | \mathrm{TS})= [ f(n-1) + f(n) + f(m) + f(m+1) ]/2$, which is the average of the lower and upper bound. This choice takes the interaction free energy in both cavities as the average of the situation where the TS particle is present or absent in the cavity. Since half of the particle in the TS is physically in contact with the particles in cavities $A$ and $B$, this is a reasonable choice. Equal importance is given to the change in free energy of both cavities, and the rate is a function of $f(n)$:
\begin{align}
k_{nm} = \frac{S n e^{- (\beta/2) \left[ f(n-1) + f(m+1) - f(n) - f(m) \right]}}{V \left( 2 \pi M \beta \right)^{1/2} e^{\beta f_{\mathrm{TS}} }}. 
\end{align}
Rewriting this as a function of $k_{10}$ gives Eq.~\eqref{eq::rates2}. Note that this rate corresponds to $c = 1/2$ in Eq.~\eqref{eq::GDBrate}.

We finally remark that the expression Eq.~\eqref{eq::rateTST} assumes that all particles crossing the TS equilibrate in cavity B. This is generally not the case: The particle can recross the TS surface on short time scales and equilibrate in cavity A. Recrossings can be accounted for by including a dynamical correction factor in the rates, which is determined from short MD simulations \cite{JCPdubbeldam2005}.
Studies of diffusion in microporous materials using dynamically corrected TST can be found in Refs.~\cite{PRLbeerdsen2004,JPCBbeerdsen2006,CESTunca2003}.

\section{Computational details}\label{app:compdet}

The computational details of the kMC simulations are given in the Supplementary Material of Ref.~\cite{PRLbecker}. In one dimension, all self-diffusions are simulated for length $L=50$. The transport diffusions are simulated for $L=20$, except for Figs.~\ref{fig::diff_fnspecial} and \ref{fig::diff_fn0_nmax2} $(L=50)$ and Fig.~\ref{fig::diff_methanol_r1} $(L=15)$. For the two- and three-dimensional systems, the concentration gradient of (labeled) particles is applied in the $x$ direction. The length in the $x$ direction is always $L_x = 15$. Periodic boundary conditions are imposed in the $y$ and $z$ directions. 
In Fig.~\ref{fig::diff_fnconcave_dimension}, the self-diffusion is simulated for $L_y = 4$ in two dimensions and $L_y = L_z = 6$ in three dimensions. The transport diffusion is measured for $L_y = 12$ in two dimensions and $L_y = L_z = 6$ in three dimensions.
In Fig.~\ref{fig::diff_fnconvex_dimension} the two-dimensional system has $L_y = 12$. For the three-dimensional system, the self-diffusion is simulated for $L_y = L_z = 6$ and the transport diffusion for $L_y = L_z = 5$.

The error bars are measured differently than in Ref.~\cite{PRLbecker}. We measure the value of the self- and transport diffusion each $10^8$ MC steps and store these values in a list $A$. Error bars are found by calculating $\overline{\sigma} = \sqrt{\left\langle \delta A^2 \right\rangle / n}$, where $n$ is the number of elements in the list and 
\begin{equation}
\left\langle \delta A^2 \right\rangle = \frac{1}{n-1} \sum_{i=1}^n (A_i - \langle A \rangle)^2,
\end{equation}
with $\langle A \rangle$ the average of the list. The error bars have value $\langle A \rangle \pm \overline{\sigma}$. The error bars are generally encompassed by the symbol sizes.


\begin{thebibliography}{66}
\expandafter\ifx\csname natexlab\endcsname\relax\def\natexlab#1{#1}\fi
\expandafter\ifx\csname bibnamefont\endcsname\relax
  \def\bibnamefont#1{#1}\fi
\expandafter\ifx\csname bibfnamefont\endcsname\relax
  \def\bibfnamefont#1{#1}\fi
\expandafter\ifx\csname citenamefont\endcsname\relax
  \def\citenamefont#1{#1}\fi
\expandafter\ifx\csname url\endcsname\relax
  \def\url#1{\texttt{#1}}\fi
\expandafter\ifx\csname urlprefix\endcsname\relax\def\urlprefix{URL }\fi
\providecommand{\bibinfo}[2]{#2}
\providecommand{\eprint}[2][]{\url{#2}}

\bibitem[{\citenamefont{K{\"a}rger et~al.}(2012)\citenamefont{K{\"a}rger,
  Ruthven, and Theodorou}}]{book_diffnano}
\bibinfo{author}{\bibfnamefont{J.}~\bibnamefont{K{\"a}rger}},
  \bibinfo{author}{\bibfnamefont{D.~M.} \bibnamefont{Ruthven}},
  \bibnamefont{and} \bibinfo{author}{\bibfnamefont{D.~N.}
  \bibnamefont{Theodorou}}, \emph{\bibinfo{title}{Diffusion in Nanoporous
  Materials}} (\bibinfo{publisher}{Wiley-VCH, New York}, \bibinfo{year}{2012}).

\bibitem[{\citenamefont{Satterfield}(1970)}]{satterfield1981mass}
\bibinfo{author}{\bibfnamefont{C.~N.} \bibnamefont{Satterfield}},
  \emph{\bibinfo{title}{Mass Transfer in Heterogeneous Catalysis}}
  (\bibinfo{publisher}{M.I.T. Press, Cambridge, MA}, \bibinfo{year}{1970}).

\bibitem[{\citenamefont{Smit and Maesen}(2008)}]{CRSmit}
\bibinfo{author}{\bibfnamefont{B.}~\bibnamefont{Smit}} \bibnamefont{and}
  \bibinfo{author}{\bibfnamefont{T.~L.~M.} \bibnamefont{Maesen}},
  \bibinfo{journal}{Chem. Rev.} \textbf{\bibinfo{volume}{108}},
  \bibinfo{pages}{4125} (\bibinfo{year}{2008}).

\bibitem[{\citenamefont{Sumida et~al.}(2012)\citenamefont{Sumida, Rogow, Mason,
  McDonald, Bloch, Herm, Bae, and Long}}]{sumida2011carbon}
\bibinfo{author}{\bibfnamefont{K.}~\bibnamefont{Sumida}},
  \bibinfo{author}{\bibfnamefont{D.~L.} \bibnamefont{Rogow}},
  \bibinfo{author}{\bibfnamefont{J.~A.} \bibnamefont{Mason}},
  \bibinfo{author}{\bibfnamefont{T.~M.} \bibnamefont{McDonald}},
  \bibinfo{author}{\bibfnamefont{E.~D.} \bibnamefont{Bloch}},
  \bibinfo{author}{\bibfnamefont{Z.~R.} \bibnamefont{Herm}},
  \bibinfo{author}{\bibfnamefont{T.-H.} \bibnamefont{Bae}}, \bibnamefont{and}
  \bibinfo{author}{\bibfnamefont{J.~R.} \bibnamefont{Long}},
  \bibinfo{journal}{Chem. Rev.} \textbf{\bibinfo{volume}{112}},
  \bibinfo{pages}{724} (\bibinfo{year}{2012}).

\bibitem[{\citenamefont{Yaghi et~al.}(2003)\citenamefont{Yaghi, O'Keeffe,
  Ockwig, Chae, Eddaoudi, and Kim}}]{Natureyaghi2003}
\bibinfo{author}{\bibfnamefont{O.~M.} \bibnamefont{Yaghi}},
  \bibinfo{author}{\bibfnamefont{M.}~\bibnamefont{O'Keeffe}},
  \bibinfo{author}{\bibfnamefont{N.~W.} \bibnamefont{Ockwig}},
  \bibinfo{author}{\bibfnamefont{H.~K.} \bibnamefont{Chae}},
  \bibinfo{author}{\bibfnamefont{M.}~\bibnamefont{Eddaoudi}}, \bibnamefont{and}
  \bibinfo{author}{\bibfnamefont{J.}~\bibnamefont{Kim}},
  \bibinfo{journal}{Nature} \textbf{\bibinfo{volume}{423}},
  \bibinfo{pages}{705} (\bibinfo{year}{2003}).

\bibitem[{\citenamefont{Davis}(2002)}]{porousapp}
\bibinfo{author}{\bibfnamefont{M.~E.} \bibnamefont{Davis}},
  \bibinfo{journal}{Nature} \textbf{\bibinfo{volume}{417}},
  \bibinfo{pages}{813} (\bibinfo{year}{2002}).

\bibitem[{\citenamefont{K{\"a}rger et~al.}(2014)\citenamefont{K{\"a}rger,
  Binder, Chmelik, Hibbe, Krautscheid, Krishna, and
  Weitkamp}}]{Naturekarger2014}
\bibinfo{author}{\bibfnamefont{J.}~\bibnamefont{K{\"a}rger}},
  \bibinfo{author}{\bibfnamefont{T.}~\bibnamefont{Binder}},
  \bibinfo{author}{\bibfnamefont{C.}~\bibnamefont{Chmelik}},
  \bibinfo{author}{\bibfnamefont{F.}~\bibnamefont{Hibbe}},
  \bibinfo{author}{\bibfnamefont{H.}~\bibnamefont{Krautscheid}},
  \bibinfo{author}{\bibfnamefont{R.}~\bibnamefont{Krishna}}, \bibnamefont{and}
  \bibinfo{author}{\bibfnamefont{J.}~\bibnamefont{Weitkamp}},
  \bibinfo{journal}{Nature Mater.} \textbf{\bibinfo{volume}{13}},
  \bibinfo{pages}{333} (\bibinfo{year}{2014}).

\bibitem[{\citenamefont{Krishna}(2012)}]{CSRkrishna2012}
\bibinfo{author}{\bibfnamefont{R.}~\bibnamefont{Krishna}},
  \bibinfo{journal}{Chem. Soc. Rev.} \textbf{\bibinfo{volume}{41}},
  \bibinfo{pages}{3099} (\bibinfo{year}{2012}).

\bibitem[{\citenamefont{Beerdsen
  et~al.}(2006{\natexlab{a}})\citenamefont{Beerdsen, Dubbeldam, and
  Smit}}]{PRLbeerdsen2006}
\bibinfo{author}{\bibfnamefont{E.}~\bibnamefont{Beerdsen}},
  \bibinfo{author}{\bibfnamefont{D.}~\bibnamefont{Dubbeldam}},
  \bibnamefont{and} \bibinfo{author}{\bibfnamefont{B.}~\bibnamefont{Smit}},
  \bibinfo{journal}{Phys. Rev. Lett.} \textbf{\bibinfo{volume}{96}},
  \bibinfo{pages}{044501} (\bibinfo{year}{2006}{\natexlab{a}}).

\bibitem[{\citenamefont{Saravanan and
  Auerbach}(1997{\natexlab{a}})}]{saravanan1997modeling}
\bibinfo{author}{\bibfnamefont{C.}~\bibnamefont{Saravanan}} \bibnamefont{and}
  \bibinfo{author}{\bibfnamefont{S.~M.} \bibnamefont{Auerbach}},
  \bibinfo{journal}{J. Chem. Phys.} \textbf{\bibinfo{volume}{107}},
  \bibinfo{pages}{8120} (\bibinfo{year}{1997}{\natexlab{a}}).

\bibitem[{\citenamefont{Saravanan and
  Auerbach}(1997{\natexlab{b}})}]{saravanan1997modeling_2}
\bibinfo{author}{\bibfnamefont{C.}~\bibnamefont{Saravanan}} \bibnamefont{and}
  \bibinfo{author}{\bibfnamefont{S.~M.} \bibnamefont{Auerbach}},
  \bibinfo{journal}{J. Chem. Phys.} \textbf{\bibinfo{volume}{107}},
  \bibinfo{pages}{8132} (\bibinfo{year}{1997}{\natexlab{b}}).

\bibitem[{\citenamefont{Demontis et~al.}(2008)\citenamefont{Demontis, Pazzona,
  and Suffritti}}]{JCPBdemontis2008}
\bibinfo{author}{\bibfnamefont{P.}~\bibnamefont{Demontis}},
  \bibinfo{author}{\bibfnamefont{F.~G.} \bibnamefont{Pazzona}},
  \bibnamefont{and} \bibinfo{author}{\bibfnamefont{G.~B.}
  \bibnamefont{Suffritti}}, \bibinfo{journal}{J. Phys. Chem. B}
  \textbf{\bibinfo{volume}{112}}, \bibinfo{pages}{12444}
  (\bibinfo{year}{2008}).

\bibitem[{\citenamefont{Auerbach}(2000)}]{IRPCauerbach}
\bibinfo{author}{\bibfnamefont{S.~M.} \bibnamefont{Auerbach}},
  \bibinfo{journal}{Int. Rev. Phys. Chem.} \textbf{\bibinfo{volume}{19}},
  \bibinfo{pages}{155} (\bibinfo{year}{2000}).

\bibitem[{\citenamefont{Reed and Ehrlich}(1981)}]{SurfScireed1981}
\bibinfo{author}{\bibfnamefont{D.~A.} \bibnamefont{Reed}} \bibnamefont{and}
  \bibinfo{author}{\bibfnamefont{G.}~\bibnamefont{Erlich}},
  \bibinfo{journal}{Surf. Sci.} \textbf{\bibinfo{volume}{102}},
  \bibinfo{pages}{588} (\bibinfo{year}{1981}).

\bibitem[{\citenamefont{Bhide and Yashonath}(1999)}]{bhide1999dependence}
\bibinfo{author}{\bibfnamefont{S.~Y.} \bibnamefont{Bhide}} \bibnamefont{and}
  \bibinfo{author}{\bibfnamefont{S.}~\bibnamefont{Yashonath}},
  \bibinfo{journal}{J. Chem. Phys.} \textbf{\bibinfo{volume}{111}},
  \bibinfo{pages}{1658} (\bibinfo{year}{1999}).

\bibitem[{\citenamefont{Zwanzig}(1992)}]{zwanzig1992diffusion}
\bibinfo{author}{\bibfnamefont{R.}~\bibnamefont{Zwanzig}}, \bibinfo{journal}{J.
  Phys. Chem.} \textbf{\bibinfo{volume}{96}}, \bibinfo{pages}{3926}
  (\bibinfo{year}{1992}).

\bibitem[{\citenamefont{Burada et~al.}(2009)\citenamefont{Burada, H{\"a}nggi,
  Marchesoni, Schmid, and Talkner}}]{Burada2009}
\bibinfo{author}{\bibfnamefont{P.~S.} \bibnamefont{Burada}},
  \bibinfo{author}{\bibfnamefont{P.}~\bibnamefont{H{\"a}nggi}},
  \bibinfo{author}{\bibfnamefont{F.}~\bibnamefont{Marchesoni}},
  \bibinfo{author}{\bibfnamefont{G.}~\bibnamefont{Schmid}}, \bibnamefont{and}
  \bibinfo{author}{\bibfnamefont{P.}~\bibnamefont{Talkner}},
  \bibinfo{journal}{ChemPhysChem} \textbf{\bibinfo{volume}{10}},
  \bibinfo{pages}{45} (\bibinfo{year}{2009}).

\bibitem[{\citenamefont{Carvalho et~al.}(2012)\citenamefont{Carvalho, Nelissen,
  Ferreira, Farias, and Peeters}}]{PREcarvalho}
\bibinfo{author}{\bibfnamefont{J.~C.~N.} \bibnamefont{Carvalho}},
  \bibinfo{author}{\bibfnamefont{K.}~\bibnamefont{Nelissen}},
  \bibinfo{author}{\bibfnamefont{W.~P.} \bibnamefont{Ferreira}},
  \bibinfo{author}{\bibfnamefont{G.~A.} \bibnamefont{Farias}},
  \bibnamefont{and} \bibinfo{author}{\bibfnamefont{F.~M.}
  \bibnamefont{Peeters}}, \bibinfo{journal}{Phys. Rev. E}
  \textbf{\bibinfo{volume}{85}}, \bibinfo{pages}{021136}
  (\bibinfo{year}{2012}).

\bibitem[{\citenamefont{Ghosh et~al.}(2010)\citenamefont{Ghosh, Marchesoni,
  Savel'ev, and Nori}}]{ghosh2010geometric}
\bibinfo{author}{\bibfnamefont{P.~K.} \bibnamefont{Ghosh}},
  \bibinfo{author}{\bibfnamefont{F.}~\bibnamefont{Marchesoni}},
  \bibinfo{author}{\bibfnamefont{S.~E.} \bibnamefont{Savel'ev}},
  \bibnamefont{and} \bibinfo{author}{\bibfnamefont{F.}~\bibnamefont{Nori}},
  \bibinfo{journal}{Phys. Rev. Lett.} \textbf{\bibinfo{volume}{104}},
  \bibinfo{pages}{020601} (\bibinfo{year}{2010}).

\bibitem[{\citenamefont{Reguera et~al.}(2012)\citenamefont{Reguera, Luque,
  Burada, Schmid, Rub\'i, and H\"anggi}}]{reguera2012entropic}
\bibinfo{author}{\bibfnamefont{D.}~\bibnamefont{Reguera}},
  \bibinfo{author}{\bibfnamefont{A.}~\bibnamefont{Luque}},
  \bibinfo{author}{\bibfnamefont{P.~S.} \bibnamefont{Burada}},
  \bibinfo{author}{\bibfnamefont{G.}~\bibnamefont{Schmid}},
  \bibinfo{author}{\bibfnamefont{J.~M.} \bibnamefont{Rub\'i}},
  \bibnamefont{and} \bibinfo{author}{\bibfnamefont{P.}~\bibnamefont{H\"anggi}},
  \bibinfo{journal}{Phys. Rev. Lett.} \textbf{\bibinfo{volume}{108}},
  \bibinfo{pages}{020604} (\bibinfo{year}{2012}).

\bibitem[{\citenamefont{Becker et~al.}(2013)\citenamefont{Becker, Nelissen,
  Cleuren, Partoens, and Van~den Broeck}}]{PRLbecker}
\bibinfo{author}{\bibfnamefont{T.}~\bibnamefont{Becker}},
  \bibinfo{author}{\bibfnamefont{K.}~\bibnamefont{Nelissen}},
  \bibinfo{author}{\bibfnamefont{B.}~\bibnamefont{Cleuren}},
  \bibinfo{author}{\bibfnamefont{B.}~\bibnamefont{Partoens}}, \bibnamefont{and}
  \bibinfo{author}{\bibfnamefont{C.}~\bibnamefont{Van~den Broeck}},
  \bibinfo{journal}{Phys.\ Rev.\ Lett.} \textbf{\bibinfo{volume}{111}},
  \bibinfo{pages}{110601} (\bibinfo{year}{2013}).

\bibitem[{\citenamefont{Chmelik et~al.}(2010)\citenamefont{Chmelik, Bux, Caro,
  Heinke, Hibbe, Titze, and K{\"a}rger}}]{PRLchmelik2010}
\bibinfo{author}{\bibfnamefont{C.}~\bibnamefont{Chmelik}},
  \bibinfo{author}{\bibfnamefont{H.}~\bibnamefont{Bux}},
  \bibinfo{author}{\bibfnamefont{J.}~\bibnamefont{Caro}},
  \bibinfo{author}{\bibfnamefont{L.}~\bibnamefont{Heinke}},
  \bibinfo{author}{\bibfnamefont{F.}~\bibnamefont{Hibbe}},
  \bibinfo{author}{\bibfnamefont{T.}~\bibnamefont{Titze}}, \bibnamefont{and}
  \bibinfo{author}{\bibfnamefont{J.}~\bibnamefont{K{\"a}rger}},
  \bibinfo{journal}{Phys.\ Rev.\ Lett.} \textbf{\bibinfo{volume}{104}},
  \bibinfo{pages}{085902} (\bibinfo{year}{2010}).

\bibitem[{\citenamefont{Krishna and van
  Baten}(2010{\natexlab{a}})}]{LANGMUIRkrishna2010_1}
\bibinfo{author}{\bibfnamefont{R.}~\bibnamefont{Krishna}} \bibnamefont{and}
  \bibinfo{author}{\bibfnamefont{J.~M.} \bibnamefont{van Baten}},
  \bibinfo{journal}{Langmuir} \textbf{\bibinfo{volume}{26}},
  \bibinfo{pages}{8450} (\bibinfo{year}{2010}{\natexlab{a}}).

\bibitem[{\citenamefont{Krishna and van
  Baten}(2010{\natexlab{b}})}]{LANGMUIRkrishnaHbond}
\bibinfo{author}{\bibfnamefont{R.}~\bibnamefont{Krishna}} \bibnamefont{and}
  \bibinfo{author}{\bibfnamefont{J.~M.} \bibnamefont{van Baten}},
  \bibinfo{journal}{Langmuir} \textbf{\bibinfo{volume}{26}},
  \bibinfo{pages}{10854} (\bibinfo{year}{2010}{\natexlab{b}}).

\bibitem[{\citenamefont{Krishna and van
  Baten}(2010{\natexlab{c}})}]{LANGMUIRkrishnaTc}
\bibinfo{author}{\bibfnamefont{R.}~\bibnamefont{Krishna}} \bibnamefont{and}
  \bibinfo{author}{\bibfnamefont{J.~M.} \bibnamefont{van Baten}},
  \bibinfo{journal}{Langmuir} \textbf{\bibinfo{volume}{26}},
  \bibinfo{pages}{3981} (\bibinfo{year}{2010}{\natexlab{c}}).

\bibitem[{\citenamefont{Esposito}(2012)}]{PREesposito2012}
\bibinfo{author}{\bibfnamefont{M.}~\bibnamefont{Esposito}},
  \bibinfo{journal}{Phys.\ Rev.\ E} \textbf{\bibinfo{volume}{85}},
  \bibinfo{pages}{041125} (\bibinfo{year}{2012}).

\bibitem[{\citenamefont{Krishna}(2009)}]{JPCCkrishna2009}
\bibinfo{author}{\bibfnamefont{R.}~\bibnamefont{Krishna}}, \bibinfo{journal}{J.
  Phys. Chem. C} \textbf{\bibinfo{volume}{113}}, \bibinfo{pages}{19756}
  (\bibinfo{year}{2009}).

\bibitem[{\citenamefont{Paschek and Krishna}(2001)}]{paschek2001}
\bibinfo{author}{\bibfnamefont{D.}~\bibnamefont{Paschek}} \bibnamefont{and}
  \bibinfo{author}{\bibfnamefont{R.}~\bibnamefont{Krishna}},
  \bibinfo{journal}{Chem. Phys. Lett.} \textbf{\bibinfo{volume}{333}},
  \bibinfo{pages}{278} (\bibinfo{year}{2001}).

\bibitem[{\citenamefont{Tunca and Ford}(2003)}]{CESTunca2003}
\bibinfo{author}{\bibfnamefont{C.}~\bibnamefont{Tunca}} \bibnamefont{and}
  \bibinfo{author}{\bibfnamefont{D.~M.} \bibnamefont{Ford}},
  \bibinfo{journal}{Chem.\ Eng.\ Sci.} \textbf{\bibinfo{volume}{58}},
  \bibinfo{pages}{3373} (\bibinfo{year}{2003}).

\bibitem[{\citenamefont{Mossa et~al.}(2004)\citenamefont{Mossa, Sciortino,
  Tartaglia, and Zaccarelli}}]{LANGMUIRmossa2004}
\bibinfo{author}{\bibfnamefont{S.}~\bibnamefont{Mossa}},
  \bibinfo{author}{\bibfnamefont{F.}~\bibnamefont{Sciortino}},
  \bibinfo{author}{\bibfnamefont{P.}~\bibnamefont{Tartaglia}},
  \bibnamefont{and}
  \bibinfo{author}{\bibfnamefont{E.}~\bibnamefont{Zaccarelli}},
  \bibinfo{journal}{Langmuir} \textbf{\bibinfo{volume}{20}},
  \bibinfo{pages}{10756} (\bibinfo{year}{2004}).

\bibitem[{\citenamefont{Ruthven}(1971)}]{NPSruthven1971}
\bibinfo{author}{\bibfnamefont{D.~M.} \bibnamefont{Ruthven}},
  \bibinfo{journal}{Nat.\ Phys.\ Sci.} \textbf{\bibinfo{volume}{232}},
  \bibinfo{pages}{70} (\bibinfo{year}{1971}).

\bibitem[{\citenamefont{Ayappa et~al.}(1999)\citenamefont{Ayappa, Kamala, and
  Abinandanan}}]{JCPayappa1}
\bibinfo{author}{\bibfnamefont{K.~G.} \bibnamefont{Ayappa}},
  \bibinfo{author}{\bibfnamefont{C.~R.} \bibnamefont{Kamala}},
  \bibnamefont{and} \bibinfo{author}{\bibfnamefont{T.~A.}
  \bibnamefont{Abinandanan}}, \bibinfo{journal}{J. Chem. Phys.}
  \textbf{\bibinfo{volume}{110}}, \bibinfo{pages}{8714} (\bibinfo{year}{1999}).

\bibitem[{\citenamefont{Ayappa}(1999)}]{JCPayappa2}
\bibinfo{author}{\bibfnamefont{K.~G.} \bibnamefont{Ayappa}},
  \bibinfo{journal}{J. Chem. Phys.} \textbf{\bibinfo{volume}{111}},
  \bibinfo{pages}{4736} (\bibinfo{year}{1999}).

\bibitem[{\citenamefont{Bae et~al.}(2008)\citenamefont{Bae, Lim, and
  Sung}}]{Langmuirbae}
\bibinfo{author}{\bibfnamefont{J.~H.} \bibnamefont{Bae}},
  \bibinfo{author}{\bibfnamefont{Y.~R.} \bibnamefont{Lim}}, \bibnamefont{and}
  \bibinfo{author}{\bibfnamefont{J.}~\bibnamefont{Sung}},
  \bibinfo{journal}{Langmuir} \textbf{\bibinfo{volume}{24}},
  \bibinfo{pages}{2569} (\bibinfo{year}{2008}).

\bibitem[{\citenamefont{Demontis et~al.}(2009)\citenamefont{Demontis, Pazzona,
  and Suffritti}}]{JCPdemontis2009}
\bibinfo{author}{\bibfnamefont{P.}~\bibnamefont{Demontis}},
  \bibinfo{author}{\bibfnamefont{F.~G.} \bibnamefont{Pazzona}},
  \bibnamefont{and} \bibinfo{author}{\bibfnamefont{G.~B.}
  \bibnamefont{Suffritti}}, \bibinfo{journal}{J.\ Chem.\ Phys.}
  \textbf{\bibinfo{volume}{130}}, \bibinfo{pages}{164701}
  (\bibinfo{year}{2009}).

\bibitem[{\citenamefont{Garc{\'\i}a et~al.}(2013)\citenamefont{Garc{\'\i}a,
  P{\'e}rez-Pellitero, Jallut, and Pirngruber}}]{LANGMUIRgarcia2013}
\bibinfo{author}{\bibfnamefont{E.~J.} \bibnamefont{Garc{\'\i}a}},
  \bibinfo{author}{\bibfnamefont{J.}~\bibnamefont{P{\'e}rez-Pellitero}},
  \bibinfo{author}{\bibfnamefont{C.}~\bibnamefont{Jallut}}, \bibnamefont{and}
  \bibinfo{author}{\bibfnamefont{G.~D.} \bibnamefont{Pirngruber}},
  \bibinfo{journal}{Langmuir} \textbf{\bibinfo{volume}{29}},
  \bibinfo{pages}{9398} (\bibinfo{year}{2013}).

\bibitem[{\citenamefont{Garc{\'\i}a et~al.}(2014)\citenamefont{Garc{\'\i}a,
  P{\'e}rez-Pellitero, Jallut, and Pirngruber}}]{JPCCgarcia}
\bibinfo{author}{\bibfnamefont{E.~J.} \bibnamefont{Garc{\'\i}a}},
  \bibinfo{author}{\bibfnamefont{J.}~\bibnamefont{P{\'e}rez-Pellitero}},
  \bibinfo{author}{\bibfnamefont{C.}~\bibnamefont{Jallut}}, \bibnamefont{and}
  \bibinfo{author}{\bibfnamefont{G.~D.} \bibnamefont{Pirngruber}},
  \bibinfo{journal}{J. Phys. Chem. C} \textbf{\bibinfo{volume}{118}},
  \bibinfo{pages}{9458} (\bibinfo{year}{2014}).

\bibitem[{\citenamefont{Van~Tassel et~al.}(1993)\citenamefont{Van~Tassel,
  Davis, and McCormick}}]{JCPvantassel}
\bibinfo{author}{\bibfnamefont{P.~R.} \bibnamefont{Van~Tassel}},
  \bibinfo{author}{\bibfnamefont{H.~T.} \bibnamefont{Davis}}, \bibnamefont{and}
  \bibinfo{author}{\bibfnamefont{A.~V.} \bibnamefont{McCormick}},
  \bibinfo{journal}{J. Chem. Phys.} \textbf{\bibinfo{volume}{98}},
  \bibinfo{pages}{8919} (\bibinfo{year}{1993}).

\bibitem[{\citenamefont{Van~Tassel et~al.}(1994)\citenamefont{Van~Tassel,
  Davis, and McCormick}}]{Langmuirvantassel}
\bibinfo{author}{\bibfnamefont{P.~R.} \bibnamefont{Van~Tassel}},
  \bibinfo{author}{\bibfnamefont{H.~T.} \bibnamefont{Davis}}, \bibnamefont{and}
  \bibinfo{author}{\bibfnamefont{A.~V.} \bibnamefont{McCormick}},
  \bibinfo{journal}{Langmuir} \textbf{\bibinfo{volume}{10}},
  \bibinfo{pages}{1257} (\bibinfo{year}{1994}).

\bibitem[{\citenamefont{Ruthven}(1984)}]{BOOKruthven}
\bibinfo{author}{\bibfnamefont{D.~M.} \bibnamefont{Ruthven}},
  \emph{\bibinfo{title}{Principles of adsorption and adsorption processes}}
  (\bibinfo{publisher}{John Wiley and Sons, Inc.}, \bibinfo{year}{1984}).

\bibitem[{\citenamefont{Gelb et~al.}(1999)\citenamefont{Gelb, Gubbins,
  Radhakrishnan, and Sliwinska-Bartkowiak}}]{RPPgelb1999}
\bibinfo{author}{\bibfnamefont{L.~D.} \bibnamefont{Gelb}},
  \bibinfo{author}{\bibfnamefont{K.~E.} \bibnamefont{Gubbins}},
  \bibinfo{author}{\bibfnamefont{R.}~\bibnamefont{Radhakrishnan}},
  \bibnamefont{and}
  \bibinfo{author}{\bibfnamefont{M.}~\bibnamefont{Sliwinska-Bartkowiak}},
  \bibinfo{journal}{Rep.\ Prog.\ Phys.} \textbf{\bibinfo{volume}{62}},
  \bibinfo{pages}{1573} (\bibinfo{year}{1999}).

\bibitem[{\citenamefont{Melnichenko et~al.}(2004)\citenamefont{Melnichenko,
  Wignall, Cole, and Frielinghaus}}]{PRE_melnichenko}
\bibinfo{author}{\bibfnamefont{Y.~B.} \bibnamefont{Melnichenko}},
  \bibinfo{author}{\bibfnamefont{G.~D.} \bibnamefont{Wignall}},
  \bibinfo{author}{\bibfnamefont{D.~R.} \bibnamefont{Cole}}, \bibnamefont{and}
  \bibinfo{author}{\bibfnamefont{H.}~\bibnamefont{Frielinghaus}},
  \bibinfo{journal}{Phys. Rev. E} \textbf{\bibinfo{volume}{69}},
  \bibinfo{pages}{057102} (\bibinfo{year}{2004}).

\bibitem[{\citenamefont{Melnichenko et~al.}(2005)\citenamefont{Melnichenko,
  Wignall, Cole, Frielinghaus, and Bulavin}}]{JML_melnichenko}
\bibinfo{author}{\bibfnamefont{Y.~B.} \bibnamefont{Melnichenko}},
  \bibinfo{author}{\bibfnamefont{G.~D.} \bibnamefont{Wignall}},
  \bibinfo{author}{\bibfnamefont{D.~R.} \bibnamefont{Cole}},
  \bibinfo{author}{\bibfnamefont{H.}~\bibnamefont{Frielinghaus}},
  \bibnamefont{and} \bibinfo{author}{\bibfnamefont{L.~A.}
  \bibnamefont{Bulavin}}, \bibinfo{journal}{J. Mol. Liq.}
  \textbf{\bibinfo{volume}{120}}, \bibinfo{pages}{7} (\bibinfo{year}{2005}).

\bibitem[{\citenamefont{H{\"a}nggi et~al.}(1990)\citenamefont{H{\"a}nggi,
  Talkner, and Borkovec}}]{RMPhanggi}
\bibinfo{author}{\bibfnamefont{P.}~\bibnamefont{H{\"a}nggi}},
  \bibinfo{author}{\bibfnamefont{P.}~\bibnamefont{Talkner}}, \bibnamefont{and}
  \bibinfo{author}{\bibfnamefont{M.}~\bibnamefont{Borkovec}},
  \bibinfo{journal}{Rev.\ Mod.\ Phys.} \textbf{\bibinfo{volume}{62}},
  \bibinfo{pages}{251} (\bibinfo{year}{1990}).

\bibitem[{\citenamefont{Ala-Nissila et~al.}(2002)\citenamefont{Ala-Nissila,
  Ferrando, and Ying}}]{ADVPHYSalanissila2002}
\bibinfo{author}{\bibfnamefont{T.}~\bibnamefont{Ala-Nissila}},
  \bibinfo{author}{\bibfnamefont{R.}~\bibnamefont{Ferrando}}, \bibnamefont{and}
  \bibinfo{author}{\bibfnamefont{S.~C.} \bibnamefont{Ying}},
  \bibinfo{journal}{Adv.\ Phys.} \textbf{\bibinfo{volume}{51}},
  \bibinfo{pages}{949} (\bibinfo{year}{2002}).

\bibitem[{\citenamefont{Evans and Hanney}(2005)}]{ZRPevansreview}
\bibinfo{author}{\bibfnamefont{M.~R.} \bibnamefont{Evans}} \bibnamefont{and}
  \bibinfo{author}{\bibfnamefont{T.}~\bibnamefont{Hanney}},
  \bibinfo{journal}{J. Phys. A: Math. Gen.} \textbf{\bibinfo{volume}{38}},
  \bibinfo{pages}{R195} (\bibinfo{year}{2005}).

\bibitem[{\citenamefont{Becker et~al.}(2014)\citenamefont{Becker, Nelissen,
  Cleuren, Partoens, and Van~den Broeck}}]{EPJSTbecker}
\bibinfo{author}{\bibfnamefont{T.}~\bibnamefont{Becker}},
  \bibinfo{author}{\bibfnamefont{K.}~\bibnamefont{Nelissen}},
  \bibinfo{author}{\bibfnamefont{B.}~\bibnamefont{Cleuren}},
  \bibinfo{author}{\bibfnamefont{B.}~\bibnamefont{Partoens}}, \bibnamefont{and}
  \bibinfo{author}{\bibfnamefont{C.}~\bibnamefont{Van~den Broeck}},
  \bibinfo{journal}{arXiv:1406.7164}  (\bibinfo{year}{2014}).

\bibitem[{\citenamefont{Vattulainen et~al.}(1999)\citenamefont{Vattulainen,
  Ying, Ala-Nissila, and Merikoski}}]{PRBvattulainen1999}
\bibinfo{author}{\bibfnamefont{I.}~\bibnamefont{Vattulainen}},
  \bibinfo{author}{\bibfnamefont{S.~C.} \bibnamefont{Ying}},
  \bibinfo{author}{\bibfnamefont{T.}~\bibnamefont{Ala-Nissila}},
  \bibnamefont{and}
  \bibinfo{author}{\bibfnamefont{J.}~\bibnamefont{Merikoski}},
  \bibinfo{journal}{Phys.\ Rev.\ B} \textbf{\bibinfo{volume}{59}},
  \bibinfo{pages}{7697} (\bibinfo{year}{1999}).

\bibitem[{\citenamefont{Taylor and Langmuir}(1933)}]{taylor1933}
\bibinfo{author}{\bibfnamefont{J.~B.} \bibnamefont{Taylor}} \bibnamefont{and}
  \bibinfo{author}{\bibfnamefont{I.}~\bibnamefont{Langmuir}},
  \bibinfo{journal}{Phys. Rev.} \textbf{\bibinfo{volume}{44}},
  \bibinfo{pages}{423} (\bibinfo{year}{1933}).

\bibitem[{\citenamefont{Nelissen et~al.}(2007)\citenamefont{Nelissen, Misko,
  and Peeters}}]{kwinten2012epl}
\bibinfo{author}{\bibfnamefont{K.}~\bibnamefont{Nelissen}},
  \bibinfo{author}{\bibfnamefont{V.~R.} \bibnamefont{Misko}}, \bibnamefont{and}
  \bibinfo{author}{\bibfnamefont{F.~M.} \bibnamefont{Peeters}},
  \bibinfo{journal}{Europhys. Lett.} \textbf{\bibinfo{volume}{80}},
  \bibinfo{pages}{56004} (\bibinfo{year}{2007}).

\bibitem[{\citenamefont{Lucena et~al.}(2012)\citenamefont{Lucena, Tkachenko,
  Nelissen, Misko, Ferreira, Farias, and Peeters}}]{kwinten2012pre85}
\bibinfo{author}{\bibfnamefont{D.}~\bibnamefont{Lucena}},
  \bibinfo{author}{\bibfnamefont{D.~V.} \bibnamefont{Tkachenko}},
  \bibinfo{author}{\bibfnamefont{K.}~\bibnamefont{Nelissen}},
  \bibinfo{author}{\bibfnamefont{V.~R.} \bibnamefont{Misko}},
  \bibinfo{author}{\bibfnamefont{W.~P.} \bibnamefont{Ferreira}},
  \bibinfo{author}{\bibfnamefont{G.~A.} \bibnamefont{Farias}},
  \bibnamefont{and} \bibinfo{author}{\bibfnamefont{F.~M.}
  \bibnamefont{Peeters}}, \bibinfo{journal}{Phys. Rev. E}
  \textbf{\bibinfo{volume}{85}}, \bibinfo{pages}{031147}
  (\bibinfo{year}{2012}).

\bibitem[{\citenamefont{Gomer}(1990)}]{RPPgomer1990}
\bibinfo{author}{\bibfnamefont{R.}~\bibnamefont{Gomer}}, \bibinfo{journal}{Rep.
  Prog. Phys.} \textbf{\bibinfo{volume}{53}}, \bibinfo{pages}{917}
  (\bibinfo{year}{1990}).

\bibitem[{\citenamefont{Beerdsen
  et~al.}(2006{\natexlab{b}})\citenamefont{Beerdsen, Dubbeldam, and
  Smit}}]{JPCBbeerdsen2006}
\bibinfo{author}{\bibfnamefont{E.}~\bibnamefont{Beerdsen}},
  \bibinfo{author}{\bibfnamefont{D.}~\bibnamefont{Dubbeldam}},
  \bibnamefont{and} \bibinfo{author}{\bibfnamefont{B.}~\bibnamefont{Smit}},
  \bibinfo{journal}{J.\ Phys.\ Chem.\ B} \textbf{\bibinfo{volume}{110}},
  \bibinfo{pages}{22754} (\bibinfo{year}{2006}{\natexlab{b}}).

\bibitem[{\citenamefont{Krishna and van Baten}(2013)}]{PCCPkrishna2013}
\bibinfo{author}{\bibfnamefont{R.}~\bibnamefont{Krishna}} \bibnamefont{and}
  \bibinfo{author}{\bibfnamefont{J.~M.} \bibnamefont{van Baten}},
  \bibinfo{journal}{Phys.\ Chem.\ Chem.\ Phys.} \textbf{\bibinfo{volume}{15}},
  \bibinfo{pages}{7994} (\bibinfo{year}{2013}).

\bibitem[{\citenamefont{Pazzona et~al.}(2009)\citenamefont{Pazzona, Demontis,
  and Suffritti}}]{JCPpazzona20092}
\bibinfo{author}{\bibfnamefont{F.~G.} \bibnamefont{Pazzona}},
  \bibinfo{author}{\bibfnamefont{P.}~\bibnamefont{Demontis}}, \bibnamefont{and}
  \bibinfo{author}{\bibfnamefont{G.~B.} \bibnamefont{Suffritti}},
  \bibinfo{journal}{J. Chem. Phys.} \textbf{\bibinfo{volume}{131}},
  \bibinfo{pages}{234704} (\bibinfo{year}{2009}).

\bibitem[{\citenamefont{Beerdsen et~al.}(2004)\citenamefont{Beerdsen, Smit, and
  Dubbeldam}}]{PRLbeerdsen2004}
\bibinfo{author}{\bibfnamefont{E.}~\bibnamefont{Beerdsen}},
  \bibinfo{author}{\bibfnamefont{B.}~\bibnamefont{Smit}}, \bibnamefont{and}
  \bibinfo{author}{\bibfnamefont{D.}~\bibnamefont{Dubbeldam}},
  \bibinfo{journal}{Phys.\ Rev.\ Lett.} \textbf{\bibinfo{volume}{93}},
  \bibinfo{pages}{248301} (\bibinfo{year}{2004}).

\bibitem[{\citenamefont{Pazzona et~al.}(2013)\citenamefont{Pazzona, Demontis,
  and Suffritti}}]{JPCCpazzona2012}
\bibinfo{author}{\bibfnamefont{F.~G.} \bibnamefont{Pazzona}},
  \bibinfo{author}{\bibfnamefont{P.}~\bibnamefont{Demontis}}, \bibnamefont{and}
  \bibinfo{author}{\bibfnamefont{G.~B.} \bibnamefont{Suffritti}},
  \bibinfo{journal}{J.\ Phys.\ Chem.\ C} \textbf{\bibinfo{volume}{117}},
  \bibinfo{pages}{349} (\bibinfo{year}{2013}).

\bibitem[{\citenamefont{Chokbunpiam et~al.}(2013)\citenamefont{Chokbunpiam,
  Chanajaree, Saengsawang, Reimann, Chmelik, Fritzsche, Caro, Remsungnen, and
  Hannongbua}}]{MMMchokbunpiam2013}
\bibinfo{author}{\bibfnamefont{T.}~\bibnamefont{Chokbunpiam}},
  \bibinfo{author}{\bibfnamefont{R.}~\bibnamefont{Chanajaree}},
  \bibinfo{author}{\bibfnamefont{O.}~\bibnamefont{Saengsawang}},
  \bibinfo{author}{\bibfnamefont{S.}~\bibnamefont{Reimann}},
  \bibinfo{author}{\bibfnamefont{C.}~\bibnamefont{Chmelik}},
  \bibinfo{author}{\bibfnamefont{S.}~\bibnamefont{Fritzsche}},
  \bibinfo{author}{\bibfnamefont{J.}~\bibnamefont{Caro}},
  \bibinfo{author}{\bibfnamefont{T.}~\bibnamefont{Remsungnen}},
  \bibnamefont{and}
  \bibinfo{author}{\bibfnamefont{S.}~\bibnamefont{Hannongbua}},
  \bibinfo{journal}{Microporous Mesoporous Mater.}
  \textbf{\bibinfo{volume}{174}}, \bibinfo{pages}{126} (\bibinfo{year}{2013}).

\bibitem[{\citenamefont{Abouelnasr and Smit}(2012)}]{PCCPabou}
\bibinfo{author}{\bibfnamefont{M.~K.~F.} \bibnamefont{Abouelnasr}}
  \bibnamefont{and} \bibinfo{author}{\bibfnamefont{B.}~\bibnamefont{Smit}},
  \bibinfo{journal}{Phys. Chem. Chem. Phys.} \textbf{\bibinfo{volume}{14}},
  \bibinfo{pages}{11600} (\bibinfo{year}{2012}).

\bibitem[{\citenamefont{Jee and Sholl}(2009)}]{JACSsholl}
\bibinfo{author}{\bibfnamefont{S.~E.} \bibnamefont{Jee}} \bibnamefont{and}
  \bibinfo{author}{\bibfnamefont{D.~S.} \bibnamefont{Sholl}},
  \bibinfo{journal}{J. Am. Chem. Soc.} \textbf{\bibinfo{volume}{131}},
  \bibinfo{pages}{7896} (\bibinfo{year}{2009}).

\bibitem[{\citenamefont{Dubbeldam et~al.}(2005)\citenamefont{Dubbeldam,
  Beerdsen, Vlugt, and Smit}}]{JCPdubbeldam2005}
\bibinfo{author}{\bibfnamefont{D.}~\bibnamefont{Dubbeldam}},
  \bibinfo{author}{\bibfnamefont{E.}~\bibnamefont{Beerdsen}},
  \bibinfo{author}{\bibfnamefont{T.~J.~H.} \bibnamefont{Vlugt}},
  \bibnamefont{and} \bibinfo{author}{\bibfnamefont{B.}~\bibnamefont{Smit}},
  \bibinfo{journal}{J.\ Chem.\ Phys.} \textbf{\bibinfo{volume}{122}},
  \bibinfo{pages}{224712} (\bibinfo{year}{2005}).

\bibitem[{\citenamefont{Ruthven and Derrah}(1972)}]{TSTruthven}
\bibinfo{author}{\bibfnamefont{D.~M.} \bibnamefont{Ruthven}} \bibnamefont{and}
  \bibinfo{author}{\bibfnamefont{R.~I.} \bibnamefont{Derrah}},
  \bibinfo{journal}{J. Chem. Soc., Faraday Trans. 1}
  \textbf{\bibinfo{volume}{68}}, \bibinfo{pages}{2332} (\bibinfo{year}{1972}).

\bibitem[{\citenamefont{K{\"a}rger et~al.}(1980)\citenamefont{K{\"a}rger,
  Pfeifer, and Haberlandt}}]{TSTkarger}
\bibinfo{author}{\bibfnamefont{J.}~\bibnamefont{K{\"a}rger}},
  \bibinfo{author}{\bibfnamefont{H.}~\bibnamefont{Pfeifer}}, \bibnamefont{and}
  \bibinfo{author}{\bibfnamefont{R.}~\bibnamefont{Haberlandt}},
  \bibinfo{journal}{J. Chem. Soc., Faraday Trans. 1}
  \textbf{\bibinfo{volume}{76}}, \bibinfo{pages}{1569} (\bibinfo{year}{1980}).

\bibitem[{\citenamefont{June et~al.}(1991)\citenamefont{June, Bell, and
  Theodorou}}]{JPCjune}
\bibinfo{author}{\bibfnamefont{R.~L.} \bibnamefont{June}},
  \bibinfo{author}{\bibfnamefont{A.~T.} \bibnamefont{Bell}}, \bibnamefont{and}
  \bibinfo{author}{\bibfnamefont{D.~N.} \bibnamefont{Theodorou}},
  \bibinfo{journal}{J. Phys. Chem.} \textbf{\bibinfo{volume}{95}},
  \bibinfo{pages}{8866} (\bibinfo{year}{1991}).

\bibitem[{\citenamefont{Tunca and Ford}(1999)}]{JCPTunca1999}
\bibinfo{author}{\bibfnamefont{C.}~\bibnamefont{Tunca}} \bibnamefont{and}
  \bibinfo{author}{\bibfnamefont{D.~M.} \bibnamefont{Ford}},
  \bibinfo{journal}{J.\ Chem.\ Phys.} \textbf{\bibinfo{volume}{111}},
  \bibinfo{pages}{2751} (\bibinfo{year}{1999}).

\bibitem[{\citenamefont{Tunca and Ford}(2002)}]{JPCBTunca2002}
\bibinfo{author}{\bibfnamefont{C.}~\bibnamefont{Tunca}} \bibnamefont{and}
  \bibinfo{author}{\bibfnamefont{D.~M.} \bibnamefont{Ford}},
  \bibinfo{journal}{J.\ Phys.\ Chem.\ B} \textbf{\bibinfo{volume}{106}},
  \bibinfo{pages}{10982} (\bibinfo{year}{2002}).

\end{thebibliography}

\end{document}